\documentclass[%
aps,prd,
amsmath,amssymb,
superscriptaddress,
12pt]{revtex4-2}

\usepackage[T1]{fontenc}
\usepackage[utf8]{inputenc}
\usepackage{amsmath,amsthm,amssymb,amsfonts,enumerate,enumitem,graphicx,xcolor,multirow}

\usepackage[pdfusetitle,bookmarksopen]{hyperref}
\usepackage{cleveref}

\def\ba#1\ea{\begin{align}#1\end{align}}
\def\be#1\ee{\begin{equation}#1\end{equation}}

\newcommand\uconn{Physics Department, University of Connecticut, Storrs, CT 06269-3046, USA}

\begin{document}
	
\title{Non-perturbative False Vacuum Decay Using Lattice Monte Carlo in Imaginary Time}
\author{Luchang Jin}
\affiliation{\uconn}
\author{Joshua Swaim}
\email{joshuatylerswaim@gmail.com}
\affiliation{\uconn}
\date{\today}

\begin{abstract}
We present a new method for calculating quantum tunneling rates using lattice Monte Carlo simulations in imaginary time. This method is designed with the goal of studying false vacuum decay non-perturbatively on the lattice. We derive a new formula, which is similar in form to Fermi's Golden Rule, which gives the decay rate in terms of an implicit decay amplitude. We then show how to calculate this implicit decay amplitude on the lattice. To deal with the suppression of the false vacuum state in the Euclidean path integral, we develop a new sampling method which combines results from multiple Monte Carlo simulations. For a simple family of one-dimensional quantum systems, we reproduce the tunneling rates calculated from the Schr\"{o}dinger equation.
\end{abstract}

\maketitle

\section{Introduction}
Some quantum field theories have a meta-stable state, called a false vacuum, which behaves somewhat like a ground state. False vacuum states occur when there is a large energy barrier separating regions of Hilbert space (see Section \ref{section_false_vacuum_decay} for a more detailed discussion). The false vacuum state eventually decays through quantum tunneling into the true ground state (the true vacuum).

False vacuum decay is relevant to many interesting field theories. Whenever there is a first-order phase transition, false vacuum decay can play an important role in the out-of-equilibrium dynamics of the field as it changes from one phase to another (see Section \ref{phase_transitions}). In theories with a first-order electroweak phase transitions, false vacuum decay in the early universe could produce detectable gravitational wave signals~\cite{Weir_GW_EW_review:2017, Hindmarsh_GW_EW_lecture_notes:2020} and be responsible for baryogenesis~\cite{Bodeker_baryogenesis_review:2020}. These kinds of decays have been the subject of non-perturbative lattice studies in the high-temperature limit, where the full quantum field theory can be reduced to a three-dimensional statistical mechanics problem~\cite{Moore_EW_bubble:2000, Gould_EW_bubble:2022}. For QCD at finite chemical potential, metastable states may be important in heavy ion collisions~\cite{Xu_meta_heavy_ion:2023}. False vacuum decay can also occur at zero temperature and density. For example, the Standard Model electroweak vacuum may be a (very long-lived) false vacuum state~\cite{Andreassen_SM_lifetime:2017, Chigusa_SM_decay:2018}. Finally, our work may be useful for studying the Schwinger mechanism (the decay of a strong field through pair production), since the Schwinger mechanism is related to false vacuum decay in the semi-classical limit~\cite{Ai_Sch_effect:2020}.

The standard way to calculate false vacuum decay rates is the semi-classical method of Callan and Coleman~\cite{Coleman_FVD:1977, Callan_FVD:1977} (which was adapted from earlier work by Langer on phase transitions~\cite{Langer:1967}). This method was extended by Linde to deal with quantum systems at finite temperature~\cite{Linde_finite_T:1980, Linde_finite_T:1981}. When the coupling constants are small, radiative corrections may be introduced order-by-order (see, for example, Ref.~\cite{Andreassen:2017, Bezuglov_two_loop_scalar:2018}). Numerical methods may also be used within this formalism to calculate the functional determinant that gives the next-to-leading-order corrections~\cite{Dunne:2005}.

For strongly-coupled theories, the normal semi-classical approximation of Callan and Coleman is not sufficient because the tree-level effective action does not provide enough information~\cite{Croon_nonpert:2021}. For example, quantum effects can create a false vacuum state even when this is not apparent in the classical Lagrangian~\cite{Weinberg_FVD_by_rad:1992}. An important example of this phenomenon (in a weakly-coupled theory) is the Standard Model electroweak vacuum~\cite{Andreassen_SM_lifetime:2017, Chigusa_SM_decay:2018}. An example of a strongly-coupled theory which could exhibit this phenomenon is QCD with a single massive quark. For a single massless quark, the ground state is degenerate, and chiral symmetry breaking causes the quark condensate to be either positive or negative. When the quark is massive, chiral symmetry is broken and the quark condensate is negative. However, there is still a false vacuum state which has a positive value for the quark condensate. Various alternatives to the semi-classical method of Callan and Coleman have been proposed~\cite{Braden:2018, Hertzberg_app_of_Braden_method:2019, Wang:2025ooq, Croon_nonpert:2021, Bai_flow_based:2024}, including a method that uses lattice Monte Carlo~\cite{Shen:2022}.

In this paper, we present a new method for calculating false vacuum decay rates using lattice Monte Carlo simulations. Lattice calculations have the advantage of being fully non-perturbative and systematically improvable. On the other hand, as discussed below, simulating false vacuum decay on the lattice is challenging. Recently, a first attempt was made to get false vacuum decay rates using the lattice~\cite{Shen:2022}. In this paper, we use an independent approach. In comparison with Ref.~\cite{Shen:2022}, our method allows us to calculate extremely small decay rates without restricting ourselves to smaller volumes. Our method also avoids making semi-classical approximations. Instead, our main source of systematic error comes from spectral reconstruction. The error from spectral reconstruction can lead to results which are off by a factor of two, although this could be systematically improved in the future by using more sophisticated methods for spectral reconstruction.

Lattice Markov Chain Monte Carlo (MCMC) calculations for false vacuum decay face several serious challenges. First, field configurations which enter the false vacuum have a much larger action then those associated with the ground state. As a result, those configurations are exponentially suppressed in the path integral. Second, Markov chain algorithms struggle to achieve ergodicity because there are multiple, well-separated regions of configuration space. Finally, even if the challenges with exponential suppression and ergodicity are overcome, the decay rate must be given in real time, whereas lattice Monte Carlo simulations are done using imaginary time.

In Section \ref{section_false_vacuum_decay}, we briefly introduce false vacuum decay. In Section \ref{section_method}, we describe our Monte Carlo method for calculating false vacuum decay rates. In Section \ref{section_results}, we describe our results for some one-dimensional single-particle systems. Finally, in Section \ref{section_future_work}, we discuss the generalization to field theory.

\section{False vacuum decay} \label{section_false_vacuum_decay}
\subsection{A simple example} \label{subsection_simple_example}
\begin{figure}
	\centering
	\includegraphics[width=\linewidth]{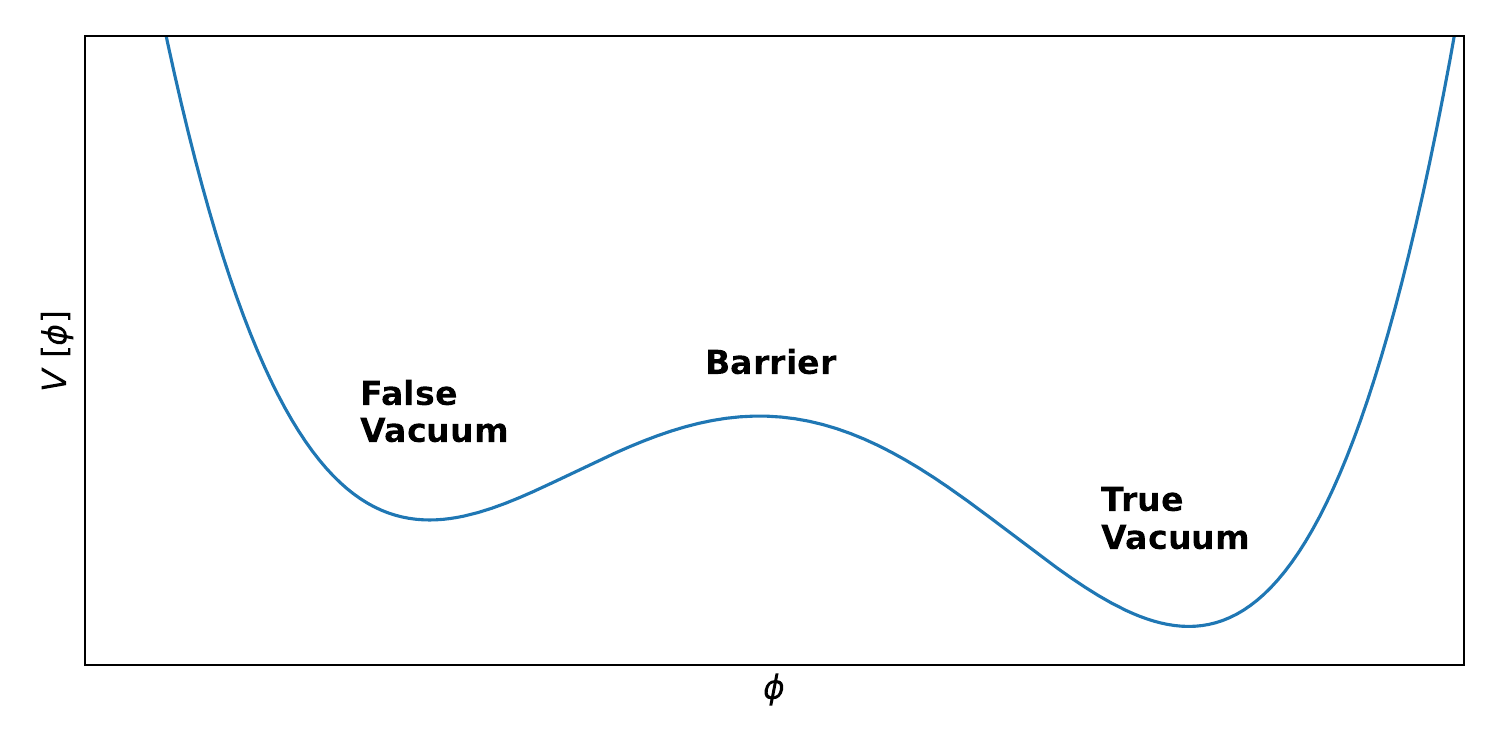}
	\caption{The potential of the field theory in Equation \ref{eq_example_Lagrangian} (for a constant field $\phi(x)=\phi$) is shown as a function of $\phi$. The regions associated with the false vacuum, the true vacuum, and the barrier between them are labeled.}
	\label{phi4_potential}
\end{figure}

A simple example of a field theory with a false vacuum state is the scalar field theory
\ba \label{eq_example_Lagrangian}
\mathcal{L}=\frac{1}{2}(\partial_\mu\phi)(\partial^\mu\phi) - V[\phi],\text{ where }V[\phi] \equiv \lambda(\phi^2-b^2)^2 - c\phi.
\ea
This theory's potential is plotted in Figure \ref{phi4_potential} for a constant field $\phi(x)=\phi$. When $c=0$, the ground state of this theory is degenerate and spontaneous symmetry breaking will occur. When $c>0$, the degeneracy is lifted and the ground state associated with $\langle \phi\rangle<0$ is given a higher energy than the true ground state. This higher-energy state is called a ``false vacuum.'' It is not an energy eigenstate and will eventually decay into the region of Hilbert space with $\langle\phi\rangle>0$.

It will be useful for us to have a more formal definition of the false vacuum state. There are multiple ways of doing this (see Appendix \ref{appendix_FGR_formal} for an alternative approach). For our purposes, we start by making a division in configuration space based on the spacial average of the configuration $\phi(x,t)$
\ba \label{eq_bar_phi}
\bar{\phi} \equiv \int d^3x \phi(\mathbf{x},t).
\ea
Configurations with $\bar\phi<0$ are associated with the false vacuum, and configurations with $\bar\phi>0$ are associated with the true vacuum. We can now construct a new quantum field theory by adding an extra term to the potential $V[\phi]$ which penalizes true vacuum configurations. We call the Hamiltonian corresponding to this new theory the ``false vacuum'' Hamiltonian $H_\text{FV}$. Note that $H$ and $H_\text{FV}$ act in the same way on any state which has no probability of being in the true vacuum region of configuration space. The ground state of $H_\text{FV}$, which we will call $|\text{FV}\rangle$, will be very close to the false vacuum state of the original theory, which we will call $|\text{FV}_\text{exact}\rangle$.

There is one point we need to deal with before finalizing our definition of $|\text{FV}_\text{exact}\rangle$. The division between the false and true vacuum regions only makes sense for low-energy states. High-energy states can move back and forth across the potential barrier with no difficulty. This accounts for the difference between $|\text{FV}\rangle$ and $|\text{FV}_\text{exact}\rangle$. The actual false vacuum state may have a tail that extends past the potential barrier, giving some ``true vacuum'' configurations a small but non-zero probability. The false vacuum Hamiltonian $H_\text{FV}$ may cut off this tail too early. This means that $|\text{FV}\rangle$ will have some overlap with high-energy eigenstates of $H$. This problem can be fixed by defining an operator which projects onto the low-energy subspace of $H$
\ba \label{eq_P_low}
P_\text{Low}\equiv \sum_{E_n<E_\text{thresh}} |E_n\rangle\langle E_n |,
\ea
where $\{|E_n\rangle\}$ are the eigenstates of $H$. $E_\text{thresh}$ must be chosen to be a little higher than the false vacuum energy. We then define the false vacuum state as
\ba \label{eq_FV_exact}
|\text{FV}_\text{exact}\rangle\equiv P_\text{Low} |\text{FV}\rangle.
\ea
Note that there is always some degree of ambiguity in any definition of the false vacuum state~\cite{Andreassen:2017}. In our definition, we could change the exact value of the energy cutoff $E_\text{thresh}$ or the exact definition of $H_\text{FV}$.

In the $\phi^4$ theory which we have just discussed, the false vacuum state is associated with an equilibrium point of the classical field theory. It should be noted that quantum effects can create a false vacuum state even when this is not apparent in the classical Lagrangian~\cite{Weinberg_FVD_by_rad:1992}, as discussed in the Introduction.

\subsection{Phase transitions and false vacuum decay} \label{phase_transitions}
There is a close analogy between false vacuum decay in quantum field theory and first-order phase transitions in statistical mechanics. In statistical mechanics, the analog of a false vacuum state is a supercooled/superheated state. This connection was pointed out by Coleman in his first paper on false vacuum decay~\cite{Coleman_FVD:1977}, and the semiclassical method of Callan and Coleman for calculating false vacuum decay rates~\cite{Callan_FVD:1977} closely followed earlier work by Langer in statistical mechanics~\cite{Langer:1967}.

False vacuum decay can occur as part of a first-order phase transition. In a supercooled/superheated state (or generalizations thereof), the unstable initial state is a pseudo-equilibrium. In a quantum system, the transition from the pseudo-equilibrium to the equilibrium phase can occur both through thermal fluctuations and through quantum tunneling (false vacuum decay). The transition rate can depend on both of these effects. The semi-classical method of Callan and Coleman was extended by Linde to deal with quantum systems at finite temperature~\cite{Linde_finite_T:1980, Linde_finite_T:1981}.

\section{Method} \label{section_method}

\subsection{Problem statement}
\begin{figure}
	\centering
	\includegraphics[width=\linewidth]{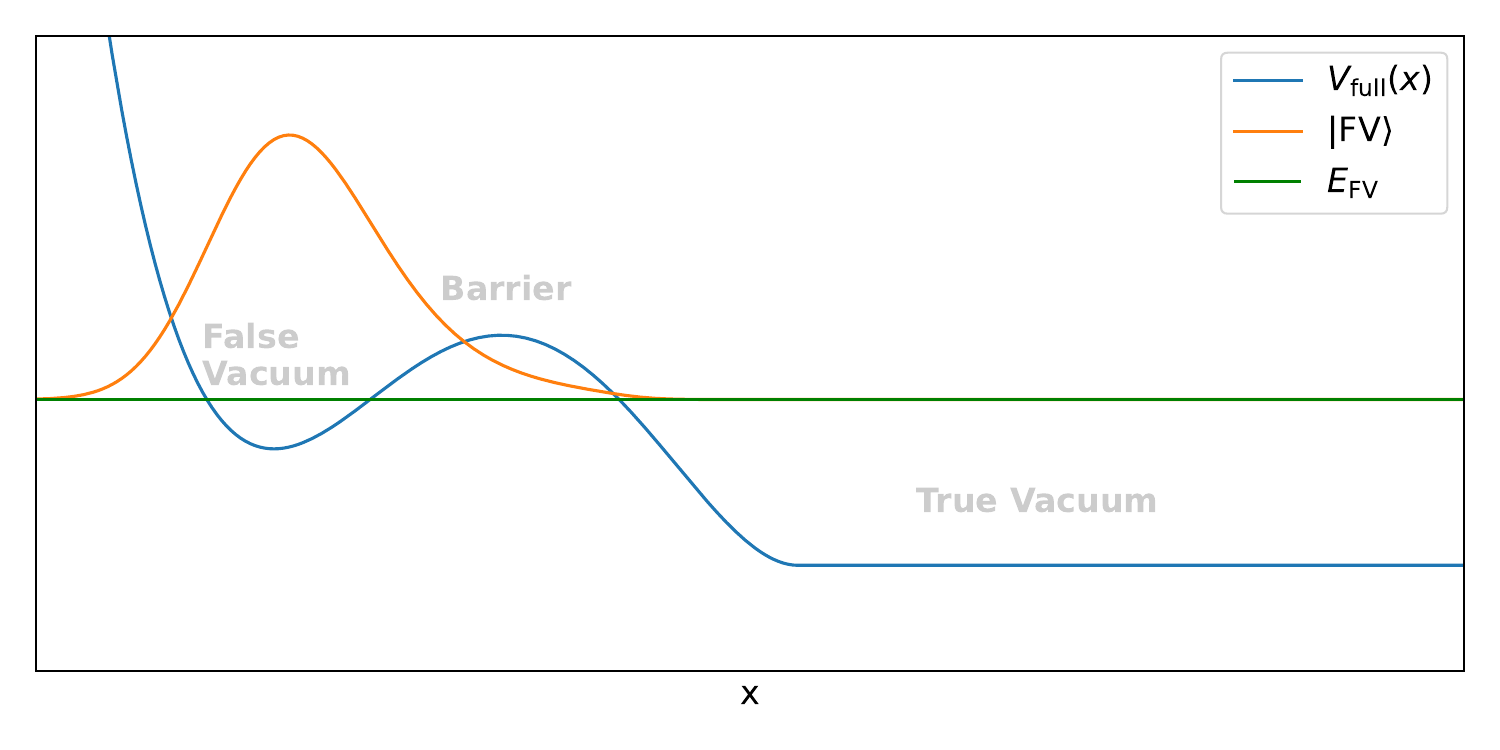}
	\caption{The potential from Equation \ref{eq_action} for a one-dimensional, single-particle quantum system is shown. We define a state $|\text{FV}\rangle$, which we will call the ``false vacuum.'' The energy of this state $E_\text{FV}$ is marked with a horizontal line, and the wavefunction for $|\text{FV}\rangle$ is plotted on top of this line (in other words, we plot $f(x)\equiv N\langle x|\text{FV}\rangle+E_\text{FV}$, where $N$ is an arbitrary normalization).}
	\label{psi_FV}
\end{figure}

Instead of dealing with a full quantum field theory, we will in this paper consider a one-dimensional, single-particle system, with an action of the form studied in Ref.~\cite{Shen:2022}
\ba \label{eq_action}
S = \beta\int dt \left[\frac{1}{2}\left(\frac{dx}{dt}\right)^2+
\begin{cases} 
	\frac{1}{2}x^2-\frac{1}{2}x^3+\frac{\alpha}{8}x^4 & x<\frac{3+\sqrt{9-8\alpha}}{2\alpha} \\
	0 & \text{otherwise}
\end{cases}\right].
\ea
The corresponding potential is plotted in Figure \ref{psi_FV} for one choice of parameters. It has two local minima separated by a potential barrier. Starting at the lower minimum and extending to the right, the potential is constant. The parameter $\alpha$ varies from 0 to 1 and controls the potential difference between between the two local minima ($\alpha=1$ gives no potential difference, while $\alpha=0$ gives infinite potential difference).

\begin{figure}
	\centering
	\includegraphics[width=\linewidth]{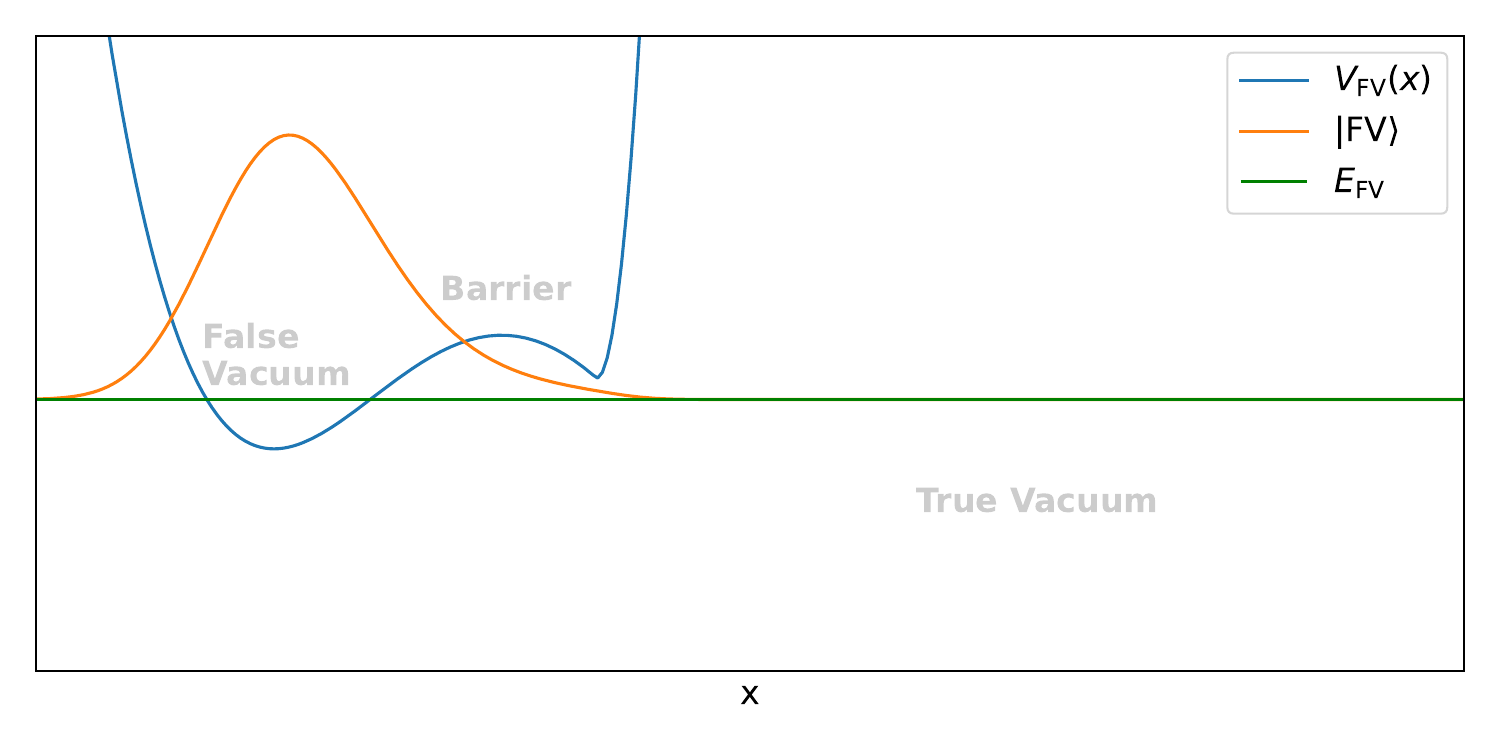}
	\caption{The false vacuum potential for the Hamiltonian from Equation \ref{eq_H_FV} is shown, along with its ground state $|\text{FV}\rangle$. The energy of this state $E_\text{FV}$ is marked with a horizontal line, and the wavefunction for $|\text{FV}\rangle$ is plotted on top of this line (in other words, we plot $f(x)\equiv N\langle x|\text{FV}\rangle+E_\text{FV}$, where $N$ is an arbitrary normalization).}
	\label{psi_FV_V_FV}
\end{figure}

As explained in Section \ref{subsection_simple_example}, the false vacuum state $|\text{FV}_\text{exact}\rangle$ is defined by projecting $|\text{FV}\rangle$ (the ground state of $H_\text{FV}$) onto the low-energy subspace of $H$. In this problem, we define $H_\text{FV}$ by
\ba \label{eq_H_FV}
H_\text{FV} \equiv H + V_\text{bar,FV},
\ea
where
\ba \label{eq_V_FV}
V_\text{bar,FV}(x)\equiv \begin{cases}
	0 & x < x_\text{bar,FV} \\ 
	-V(x) + V(x_\text{bar,FV}) + B\cdot(x-x_\text{bar,FV})^2 & \text{otherwise,}
\end{cases}
\ea
with $x_\text{bar,FV}=\frac{3+\sqrt{9-8\alpha}}{2\alpha} + x_\text{FV offset}$ lying to the right (by amount $x_\text{FV offset}$) of the maximum of the potential barrier between the false vacuum and true vacuum. The potential $V_\text{FV}$ of this Hamiltonian is plotted in Figure \ref{psi_FV_V_FV}.

Each energy eigenstate of $H$ (with energy less than the height of the potential barrier) will be exponentially suppressed in the false vacuum region unless its energy lies within a narrow band around one of the resonant energies (see, for example, Ref.~\cite{Andreassen:2017}). States which are confined to the false vacuum region will be mainly composed of eigenstates within one of these resonant bands. The false vacuum state will be be composed of energy eigenstates from the lowest resonant band. States built from the other resonant bands will tunnel through the barrier more quickly, and so the decay rate of our false vacuum state will determine the decay rate of any low-energy state initially confined to the false vacuum region in the long-time limit.

We are interested in the rate of tunneling through the potential barrier (which we will call the decay rate). The probability of remaining in the false vacuum after time $t$ is 
\ba 
P_\text{FV}(t) \equiv |\langle \text{FV}_\text{exact}|e^{-iHt}|\text{FV}_\text{exact}\rangle|^2.
\ea
We expect exponential decay $P_\text{FV}(t) = e^{-\Gamma t}$. However, this formula does not hold near $t=0$ or when $t$ is large. Nevertheless, for some range of intermediate times $t$, the decay is exponential (see, for example, Ref.~\cite{Andreassen:2017}). If we define
\ba \label{eq_decay_rate}
\Gamma(t) = -\frac{1}{P_\text{FV}}\frac{dP_\text{FV}}{dt},
\ea 
then $\Gamma(t)$ will give us a roughly constant decay rate for this intermediate range of times.

\subsection{Euclidean-time observables} \label{section_FGR}
So far, we have expressed the decay rate in terms of Minkowski-time observables. If $\Gamma(t)$ were constant all the way to $t=0$, then the analytic continuation to Euclidean time could be done trivially. Unfortunately, this is not the case. However, the fact that $\Gamma(t)$ is constant for a range of intermediate times will make the inverse problem easier. We need an observable which will predict the behavior of $\Gamma(t)$ at intermediate times from its behavior at short times. This will allow us to construct a constant observable equal to the decay rate which can be computed in both Minkowski and Euclidean time. We will draw inspiration from Fermi's golden rule:
\ba \label{eq_FGR_original}
\Gamma = 2\pi\langle \text{FV}_\text{exact}|H_\text{tr} \delta(H_0-E_\text{FV}) H_\text{tr}|\text{FV}_\text{exact}\rangle.
\ea 
Here we have defined $H_0$ and $H_\text{tr}$ such that $H=H_0+H_\text{tr}$, with $H_0$ being the part of the Hamiltonian which does not allow low-energy states to transition between the false vacuum and the true vacuum. Then $H_\text{tr}$ is the part of the Hamiltonain which does allow low-energy transitions (tunnelling). The state $H_\text{tr}|\text{FV}_\text{exact}\rangle$ depends only on the short-time time evolution of $|\text{FV}_\text{exact}\rangle$ in both Euclidean and Minkowski time.

We construct an explicit form for $H_0$ and $H_\text{tr}$ in Appendix \ref{appendix_FGR_formal}, but this approach is not very useful for practical calculations. Instead, our strategy will be to construct the transition amplitude in Equation \ref{eq_FGR_original} implicitly. We will replace $H_\text{tr}=H-H_0$ with $H-H_\text{FV}$. Similarly, we will replace $H_0$ in $\delta(H_0-E_\text{FV})$ with
\ba \label{eq_H_TV}
H_\text{TV}=H + V_\text{bar,TV},
\ea
where
\ba \label{eq_V_TV}
V_\text{bar,TV}\equiv \begin{cases}
	-V + V(x_\text{bar,TV}) + B(x-x_\text{bar,TV})^2 & x < x_\text{bar,TV} \\ 
	0 & \text{otherwise}
\end{cases}
\ea
with $x_\text{bar,TV}=\frac{3+\sqrt{9-8\alpha}}{2\alpha}$ being the location of the maximum of the potential barrier between the false vacuum and true vacuum. $H_\text{TV}$ is similar to $H$ except for an additional potential barrier that prevents states from entering the false vacuum.

\subsection{The Implicit Decay Amplitude Method} \label{section_FGR_derivation}

We will derive an equation similar to Fermi's golden rule, except we will not use $H_0$ or $H_\text{tr}$ explicitly. Instead, we will rely on $H_\text{FV}$ and $H_\text{TV}$ to play the role of $H_0$. To start, we want to calculate the transition amplitude
\ba \label{eq_decay_amp}
d_f(t) = e^{iE_\text{FV}t}\langle f|e^{-iHt}|\text{FV}_\text{exact}\rangle \equiv e^{iE_\text{FV}t}\langle f|e^{-iHt}P_\text{Low}|\text{FV}\rangle.
\ea 
Recall from Section \ref{subsection_simple_example} that $|\text{FV}\rangle$ is defined as the the ground state of $H_\text{FV}$. The exact false vacuum state $|\text{FV}_\text{exact}\rangle$ was defined in Equation \ref{eq_FV_exact} as $P_\text{Low}|\text{FV}\rangle$, where $P_\text{Low}$ projects onto the low-energy subspace of $H$. The false vacuum energy $E_\text{FV}$ is defined so that 
\ba \label{eq_E_FV}
H_\text{FV}|\text{FV}\rangle=E_\text{FV}|\text{FV}\rangle.
\ea 
We define $\langle f|$ to be an eigenstate of $H_\text{TV}$ with eigenvalue $E_f$. The phase factor $e^{iE_\text{FV}t}$ is added for later convenience.

If we take the derivative of the transition amplitude, we get
\ba\label{eq_exact_amp}
	\frac{d}{dt}d_f(t) = i\langle f|e^{-i(H-E_\text{FV})t}(E_\text{FV} - H)P_\text{Low}|\text{FV}\rangle = i\langle f|e^{-i(H-E_\text{FV})t}P_\text{Low}|D\rangle,
\ea
where we have defined the difference state
\ba \label{eq_D}
|D\rangle \equiv (H_\text{FV} - H)|\text{FV}\rangle.
\ea
In deriving Equation \ref{eq_exact_amp}, we have used the fact that $P_\text{Low}$ commutes with $H$. Now note that the state $|D\rangle$ will be non-zero only in the true vacuum (where $H_\text{FV}$ and $H$ differ). Since the initial state $|\text{FV}\rangle$ is localized in the false vacuum, with only a small tail extending out into the true vacuum, the state $|D\rangle$ will be localized around the region where $H_\text{FV}$ first begins to differ from $H$. We will call this region the false vacuum boundary. Note that our definition of $H_\text{FV}$ puts the false vacuum boundary on the right side of the potential barrier separating the false vacuum from the true vacuum (see Figure \ref{psi_FV_V_FV}).

So far, Equation \ref{eq_exact_amp} is an exact statement. Now we come to the first approximation. We can write $\langle f|e^{-i(H-E_\text{FV})t}$ as
\ba 
\langle f|e^{-i(H-E_\text{FV})t} = \langle f|e^{-i(H_\text{TV}-E_\text{FV})t} + \langle f|\left(e^{-i(H-E_\text{FV})t}-e^{-i(H_\text{TV}-E_\text{FV})}\right).
\ea
Because $H$ acts similarly to $H_\text{TV}$ on $|f\rangle$, we can approximate
\ba  \label{eq_FGR_approx}
\langle f|e^{-i(H-E_\text{FV})t} \to \langle f|e^{-i(H_\text{TV}-E_\text{FV})t}=\langle f|e^{-i(E_f-E_\text{FV})t}.
\ea
This is valid as long as
\ba \label{eq_FGR_inequality_condition}
|\langle f|P_\text{Low}|D\rangle| \gg \left|\langle f|\left(e^{-i(H-E_\text{FV})t}-e^{-i(H_\text{TV}-E_\text{FV})t}\right)P_\text{Low}|D\rangle\right|.
\ea 
Let us consider the conditions for this inequality to hold. The state $|D\rangle$ is localized on the false vacuum boundary. We can choose $H_\text{TV}$ so that it is the same as $H$ in this region. Then the only reason evolution with $H$ will differ from evolution with $H_\text{TV}$ is if the time evolution with $H$ propagates $|D\rangle$ back through the barrier into the false vacuum. For low-energy states, this is a higher-order effect (the probability of passing through the barrier twice is lower than the probability of passing through it once). It should be noted that the low-energy projection $P_\text{Low}$ is important for making this argument, since higher-energy components of $|D\rangle$ could propagate back over the barrier more easily.

With the approximation in Equation \ref{eq_FGR_approx}, Equation \ref{eq_exact_amp} becomes
\ba \label{eq_approx_amp}
\frac{d}{dt}d_f(t) \approx  ie^{-i\omega_ft}\langle f|P_\text{Low}|D\rangle
\ea 
where $\omega_f \equiv E_f-E_\text{FV}$. For convinience, we will define
\ba
|D'\rangle\equiv P_\text{Low}|D\rangle.
\ea
We can now integrate Equation \ref{eq_approx_amp} to get
\ba 
d_f(t) \approx d_f(0) + i\langle f|D'\rangle\int_0^t dt e^{-i\omega_f t} = \langle f|\text{FV}_\text{exact}\rangle + \langle f|D'\rangle\frac{1-e^{-i\omega_f t}}{\omega_f}.
\ea
We can approximate Equation \ref{eq_decay_rate} by
\ba 
\Gamma\approx -\frac{dP_\text{FV}}{dt}
\ea 
as long as we are dealing with sufficiently short times (so that $P_\text{FV}\approx 1$). Then the first-order transition rate is then
\ba 
\Gamma\approx \frac{d}{dt}|d_f(t)|^2 &\approx 2\text{Re}\left(\langle \text{FV}_\text{exact}|f\rangle\langle f|D'\rangle (ie^{-i\omega_{f}t})\right) + \left|\langle f|D'\rangle\right|^2 \frac{2\sin(\omega_{f}t)}{\omega_{f}}
\ea
Integrating over all final states, we get the total decay probability
\ba 
\int df \frac{d}{dt}|d_f(t)|^2 \approx 2\int df \left[\text{Re}\left(i\langle \text{FV}_\text{exact}|f\rangle\langle f|D'\rangle e^{-i\omega_ft}\right) + \left|\langle f|D'\rangle\right|^2 \frac{\sin(\omega_{f}t)}{\omega_{f}}\right]
\ea 
For $t\gg 1/\omega_f$, the function $e^{-i\omega_ft}$ is rapidly oscillating. Therefore, the first term integrates to zero in this limit. On the other hand, when $t\gg 1/\omega_f$, the function $\sin(\omega_{f}t)/\omega_{f}$ is sharply peaked around $\omega_{f}=0$ and integrates to $\pi$ (with respect to $E_f$). Therefore, the first-order decay rate in the limit $t\gg 1/\omega_f$ is
\ba 
\Gamma\equiv \int df \frac{d}{dt}|d_f(t)|^2  \approx 2\pi \langle D|P_\text{Low} \delta(H_\text{TV}-E_\text{FV}) P_\text{Low}|D\rangle.
\ea
Now since the delta function $\delta(H_\text{TV}-E_\text{FV})$ ensures we are in the low-energy subspace of $H_\text{TV}$, which is identical to the low-energy subspace of $H$ for states localized in the true vacuum, we can remove the low-energy projectors $P_\text{Low}$. Therefore, our final formula for the decay rate in terms of the implicit decay amplitude is
\ba \label{eq_FGR}
\Gamma\equiv 2\pi \langle \text{FV}|(H_\text{FV}-H)\delta(H_\text{TV}-E_\text{FV})(H_\text{FV}-H)|\text{FV}\rangle.
\ea
Remember that we are required to choose $H_\text{FV}$ so that it begins to differ from $H$ only on the true vacuum side of the potential barrier that separates the false vacuum and true vacuum (in other words, we require $x_\text{FV offset}>0$ in Equation \ref{eq_V_FV}). Additionally, $H_\text{TV}$ must be chosen so that it is identical to $H$ in the region where $H_\text{FV}$ and $H$ start to differ. These two conditions were required to satisfy Equation \ref{eq_FGR_inequality_condition}.

\subsection{The inverse problem}\label{section_inverse_prob}
We now turn to the problem of implementing the energy-conserving delta function in the implicit decay amplitude formula (Eq. \ref{eq_FGR}). While we cannot calculate $\langle D|\delta(H_\text{TV}-E_\text{FV})|D\rangle$ directly, we can instead calculate
\ba 
Q(t)\equiv \langle D|e^{-(H_\text{TV}-E_\text{FV})t}|D\rangle.
\ea
If we had perfect data for all $t$ without any statistical noise, we could use $Q(t)$ to infer the spectrum
\ba \label{eq_rho}
\rho(E)\equiv \langle D|\delta(H_\text{TV}-E)e^{-(H_\text{TV}-E_\text{FV})t}|D\rangle.
\ea
Note that $\rho(E_\text{FV})$ gives us the matrix element we need for the implicit decay amplitude in Equation \ref{eq_FGR}. Unfortunately, we can only calculate $Q(t)$ a finite number of times, and all our data will have statistical noise. The problem of inferring $\rho(E)$ based on $Q(t)$ is an ill-posed inverse problem.

While many methods have been proposed for dealing with this inverse problem, we will in this work simply assume $\rho(E)$ has a Gaussian distribution and use $Q(t)$ to fit for the center and width of the distribution. This ansatz will give good results as long as $\rho(E)$ has a single (roughly Gaussian) peak centered on $E_\text{FV}$ (remember that we only need $\rho(E_\text{FV})$ to be determined accurately). Because $\rho(E)$ is initially suppressed for energies below $E_\text{FV}$, suppressing high-energy states by Euclidean time evolution with $e^{-(H_\text{TV}-E_\text{FV})t}$ will often lead to a single-peak structure centered on $E_\text{FV}$ for intermediate $t$ (if $t$ is too long, we will overcome the initial suppression of the low-energy states and project out the true ground state). This is similar to how the canonical ensemble is very close to the microcanonical ensemble (except at very low temperatures). The Boltzman weight $e^{-\beta E}$ suppresses high energy states, but the phase space density is small for low-energy states, leading to a sharply-peaked energy distribution. If our Gaussian ansatz is not justified, then the fitting procedure will not give good results, and we will know that a more sophisticated method is required. On the other hand, if this ansatz is well-justified, then the fitting procedure will work very well, and we can have confidence in our results.

There are many more sophisticated methods available for spectral reconstruction which could be used in future studies. One example is the method of Hansen, Lupo, and Tantalo (HLT)~\cite{Hansen:2019idp}, which has been applied, for example, in reconstructing the $R$-ratio~\cite{Blum:2024hyr,ExtendedTwistedMass:2025tpc}. A similar method has been proposed which uses Chebyshev polynomials~\cite{Bailas:2020qmv}. Another newly introduced method is Nevanlinna-Pick interpolation~\cite{Bergamaschi:2023xzx}. In this method, it can be difficult to deal with statistical errors in the data, although there has been recent work on addressing this problem~\cite{Fields:2025glg}. Other methods include the maximum entropy method~\cite{Rietsch:1977,Asakawa:2000tr}, Bayesian reconstruction, such as in Ref. ~\cite{Burnier:2013nla}, the Bakus-Gilbert method~\cite{Backus:1968svk,Backus:1970,Hansen:2017mnd}, and newly-introduced moment problem methods~\cite{Abbott:2025snz}.

\subsection{Lattice observables} \label{section_lattice_obs}
On the lattice we will calculate
\ba 
Q_\text{lat}(t,t_\text{FV}) \equiv \frac{\text{Tr}\left[\frac{1}{a}\left(e^{-Ha}-e^{-H_\text{FV}a}\right)e^{-H_\text{TV} t}\frac{1}{a}\left(e^{-Ha}-e^{-H_\text{FV}a}\right)e^{-H_\text{FV}t_\text{FV}}\right]}{\text{Tr}\left[e^{-H_\text{FV}(2a+t+t_\text{FV})}\right]}.
\ea 
Since $|\text{FV}\rangle$ is the ground state of $H_\text{FV}$, for large enough $t_\text{FV}$, we can approximate
\ba 
Q_\text{lat}(t,t_\text{FV})\approx \langle D_\text{lat}|e^{-(H_\text{TV}-E_\text{FV}) t}|D_\text{lat}\rangle,
\ea
where
\ba 
|D_\text{lat}\rangle\equiv \frac{1}{a}\left(e^{-(H-E_\text{FV})a}-e^{-(H_\text{FV}-E_\text{FV})a}\right)|\text{FV}\rangle\approx |D\rangle\equiv (H_\text{FV}-H)|\text{FV}\rangle.
\ea

Now recall that $H_\text{FV}=H+V_\text{bar,FV}$. Therefore, in the lattice discretization, using $K$ to represent the kinetic term in the Hamiltonians,
\ba 
e^{-Ha}-e^{-H_\text{FV}a} &= e^{-Ka/2}e^{-Va}e^{-Ka/2}-e^{-Ka/2}e^{-V_\text{FV}a}e^{-Ka/2} 
\\
&=e^{-Ka/2}(e^{-Va}-e^{-V_\text{FV}a})e^{-Ka/2}
\\
&=e^{-Ka/2}e^{-Va}(1-e^{-V_\text{bar,FV}a})e^{-Ka/2}.
\ea
Because $V_\text{bar,FV}$ is positive semi-definite, all the eigenvalues of $e^{-V_\text{bar,FV}a}$ are less than or equal to 1. Therefore, $1-e^{-V_\text{bar,FV}a}$ is positive semi-definite, and $1-e^{-(V_\text{bar,FV}+\epsilon)a}$ is positive definite for any small $\epsilon>0$. Therefore, we can write,
\ba 
e^{-Ha} - e^{-H_\text{FV}a} \approx e^{-Ka/2}\exp\left\{-\big[V - \log(1-e^{-(V_\text{bar,FV}+\epsilon)a})/a\big]a\right\}e^{-Ka/2}.
\ea
If we define
\ba 
H_\text{proj}' \equiv  H - \log[1-e^{-(V_\text{bar}+\epsilon)a}]/a,
\ea
we can replace the difference $e^{-Ha}-e^{-H_\text{FV}a}$ with $e^{-H_\text{proj}'a}$. To account for the factor of $a^2$ in the denominator of $Q_\text{lat}$, we could add $\log(a)/a$ to our definition to get  
\ba \label{eq_def_V_proj}
H_\text{proj} \equiv H + V_\text{proj} \equiv  H - \log\left[\frac{1-e^{-(V_\text{bar,FV}+\epsilon)a}}{a}\right]/a.
\ea
Then
\ba 
Q_\text{lat}(t,t_\text{FV})\approx \frac{\text{Tr}\Big[e^{-H_\text{proj}a}e^{-H_\text{TV}t}e^{-H_\text{proj}a}e^{-H_\text{FV}t_\text{FV}}\Big]}{\text{Tr}\Big[e^{-H_\text{FV}(2a+t+t_\text{FV})}\Big]}.
\ea

As we will see in the next section, calculating $Q_\text{lat}(t,t_\text{FV})$ can be expensive. As we saw earlier, we need to calculate $Q_\text{lat}$ for several different times $t$ in order to solve the inverse problem. We can save computational expense by instead calculating the ratio
\ba \label{eq_Q_div_Q}
\frac{Q_\text{lat}(t, t_\text{FV})}{Q_\text{lat}(t-a, t_\text{FV}+a)}\approx \frac{\text{Tr}\Big[e^{-H_\text{proj}a}e^{-H_\text{TV}t}e^{-H_\text{proj}a}e^{-H_\text{FV}t_\text{FV}}\Big]}{\text{Tr}\Big[e^{-H_\text{proj}a}e^{-H_\text{TV}(t-a)}e^{-H_\text{proj}a}e^{-H_\text{FV}(t_\text{FV}+a)}\Big]}.
\ea
As we will see in the next section, this ratio is much less expensive to calculate. Calculating this ratio at several different points in time allows us to fit for the spectrum $\rho(E)$ without calculating $Q$ itself multiple times.

\subsection{The Intermediate Ratios Method} \label{section_intermediate_ratios_method}

We now turn to the problem of calculating $Q_\text{lat}(t,t_\text{FV})$ and $Q_\text{lat}(t,t_\text{FV})/Q_\text{lat}(t-a,t_\text{FV}+a)$ on the lattice. We want to calculate ratios of the form
\ba 
\frac{\text{Tr}\left[\prod_i \exp({-H_{N,i}t_i})\right]}{\text{Tr}\left[\prod_i \exp({-H_{D,i}t_i})\right]} = \frac{\int\mathcal{D}x \exp(-S_N)}{\int\mathcal{D}x \exp({-S_D})},
\ea 
where
\ba 
S_j[x] \equiv \sum_i\int_{t_{i-1}}^{t_i} dt H_{j,i}(x(t),\dot{x}(t)).
\ea
Ratios of traces of the above form can be calculated using Monte Carlo simulations. If we define
\ba 
\Delta S[x] = S_N-S_D
\ea 
then
\ba 
\frac{\int\mathcal{D}x \exp({-S_N})}{\int\mathcal{D}x \exp({-S_D})}=\frac{\int\mathcal{D}x \exp(-S_D)\exp(-\Delta S)}{\int\mathcal{D}x \exp({-S_D})}.
\ea 

Unfortunately, $\exp({-\Delta S})$ may get exponentially large contributions from configurations that are exponentially suppressed in the path integral by $\exp({-S_D})$. To avoid this problem, we must ensure that we get similar ensembles whether we sample configurations according to $\frac{1}{Z_D}\mathcal{D}xe^{-S_D[x]}$ or according to $\frac{1}{Z_N}\mathcal{D}xe^{-S_N[x]}$. In particular, we want to avoid a situation where $e^{-S_D}$ suppresses configurations which are important for $e^{-S_N}$. We can ensure this by imposing conditions on the relationship between $S_D$ and $S_N$.
\begin{enumerate}
	\item We generally want $S_D[x]\leq S_N[x]$ for all configurations $x$. Otherwise, configurations which are exponentially important to the observable $\exp(-(S_N-S_D))$ might be exponentially suppressed by $\exp(-S_D)$ in the path integral.
	\item We want $S_D[x]$ and $S_N[x]$ to be similar to each other.
\end{enumerate}

The actions defined by the numerator and denominator of $Q_\text{lat}$ and $Q_\text{lat}(t+a)/Q_\text{lat}(t)$ do not obey these conditions ($Q_\text{lat}$ does not obey either one, while $Q_\text{lat}(t+a)/Q_\text{lat}(t)$ obeys only the second condition). However, we can fix this problem by defining multiple intermediate actions $S_i$ which interpolate smoothly between $S_D$ and $S_N$.

We can define two families of intermediate actions obeying both conditions which together interpolate between $S_D$ and $S_N$
\ba 
S_L(L) \equiv \sum_i\int_{t_{i-1}}^{t_i} dt \Big[H_{D,i} + L\max(0, H_{N,i}-H_{D_i}) \Big]
\ea 
and
\ba 
S_M(M) \equiv \sum_i\int_{t_{i-1}}^{t_i} dt \Big[H_{N,i} + M\max(0, H_{D,i}-H_{N_i}) \Big].
\ea 
These are defined so that $S_L(L_1)[x]\leq S_L(L_2)[x]$ whenever $L_1<L_2$ and $S_M(M_1)[x]\leq S_M(M_2)[x]$ whenever $M_1<M_2$. Note also that
\ba 
S_L(0)=S_D,\quad S_M(0)=S_N,\quad \text{and}\quad S_L(1)=S_M(1)=\sum_i\int_{t_{i-1}}^{t_i} dt \max(H_{N_i}, H_{D,i}).
\ea 
We can calculate the ratio we want using
\ba 
\frac{\int\mathcal{D}x \exp({-S_N})}{\int\mathcal{D}x \exp({-S_D})} = \prod_{j=0}^l \frac{\int\mathcal{D}x \exp({-S_L(L_{j+1})})}{\int\mathcal{D}x \exp({-S_L(L_{j})})} \left(\prod_{k=0}^m \frac{\int\mathcal{D}x \exp({-S_M(M_{k+1})})}{\int\mathcal{D}x \exp({-S_M(M_{k})})}\right)^{-1},
\ea 
where $L_0=M_0=0$, $L_l=M_m=1$, and the sequences $(L_j)$ and $(M_k)$ are strictly increasing. Note that we can calculate an observable like
\ba 
{\int\mathcal{D}x \exp({-S_M(M_{j+1})})}/{\int\mathcal{D}x \exp({-S_M(M_{j})})}
\ea 
using a Monte Carlo simulation before inverting it to get its contribution to $Q_\text{lat}$. In this way, we can always choose the smaller action to be in the denominator.

\subsubsection{Avoiding Issues with Ergodicity}
In defining the intermediate actions, it is important to avoid issues with ergodicity. Note that the numerator and denominator actions for both $Q_\text{lat}$ and $Q_\text{lat}(t, t_\text{FV})/Q_\text{lat}(t-a, t_\text{FV}+a)$ do not have any problems with ergodicity because the potentials at each point in time have only a single local minimum ($H_\text{FV}$ technically has two minima, but the second one is much higher by construction and therefore is not important except near $H_\text{proj}$).

If necessary, we can define two new families of intermediate potentials. The first interpolates between $S_D$ and $S_D'$, and the second interpolates between $S_N$ and $S_N'$, where we choose $S_D'$ and $S_N'$ to avoid issues with ergodicity in defining $S_L$ and $S_M$. There are many ways of doing this. For example, we could define $S_D'=C S_D$ and $S_N'= CS_N$, where $0<C<1$. By the analogy between quantum and statistical mechanics, this would correspond to ``raising the temperature.''

In our calculation, we take $S_N'=S_N$, since no ergodicity issues arise. For $Q_\text{lat}(t, t_\text{FV})$ / $Q_\text{lat}(t-a, t_\text{FV}+a)$, we also take $S_D'=S_D$. However, when calculating $Q_\text{lat}(t, t_\text{FV})$, we define a family of actions $S_D(P)$ with $S_D(0)=S_D$ and $S_D'=S_D(1)$. $S_D(P)$ is defined to be equal to $S_D$ except that we replace $V_\text{FV}$ on the time slices where $V_\text{proj}$ occurs in $S_N$ with 
\ba \label{eq_S_P}
V_\text{FV} \to \max[V_\text{FV}(x), (P-1)V_\text{FV}(x_\text{min}) + PV_\text{FV}(x_\text{bar})],
\ea 
where $x_\text{min}$ is the location of the false vacuum local minimum (the global minimum of $V_\text{FV}$). We then modify our definition of $S_L(L)$ and $S_M(M)$ to start from $S_D'$ and $S_N'$ instead of $S_D$ and $S_N$:
\ba \label{eq_S_L}
S_L(L) \equiv \sum_i\int_{t_{i-1}}^{t_i} dt \Big[H_{D,i}' + L\max(0, H_{N,i}'-H_{D_i}') \Big]
\ea 
and
\ba \label{eq_S_M}
S_M(M) \equiv \sum_i\int_{t_{i-1}}^{t_i} dt \Big[H_{N,i}' + M\max(0, H_{D,i}'-H_{N_i}') \Big].
\ea 

\subsubsection{Calculating the Observables} \label{section_intermediate_ratios_calc_obs}
We can calculate $Q_\text{lat}$ using
\ba 
Q_\text{lat} = \frac{\int\mathcal{D}x \exp({-S_N})}{\int\mathcal{D}x \exp({-S_D})} = \prod_{i=0}^p \frac{\int\mathcal{D}x \exp({-S_D(P_{i+1})})}{\int\mathcal{D}x \exp({-S_D(P_{i})})}\left(\frac{\int\mathcal{D}x \exp({-S_N})}{\int\mathcal{D}x \exp({-S_D}')}\right)
\ea
\ba
= \prod_{i=0}^p \frac{\int\mathcal{D}x \exp({-S_D(P_{i+1})})}{\int\mathcal{D}x \exp({-S_D(P_{i})})}\prod_{j=0}^l \frac{\int\mathcal{D}x \exp({-S_L(L_{j+1})})}{\int\mathcal{D}x \exp({-S_L(L_{j})})} \left(\prod_{k=0}^m \frac{\int\mathcal{D}x \exp({-S_M(M_{k+1})})}{\int\mathcal{D}x \exp({-S_M(M_{k})})}\right)^{-1},
\ea 
where $P_0=L_0=M_0=0$, $P_p=L_l=M_m=1$, and the sequences $(P_i)$, $(L_j)$, and $(M_k)$ are strictly increasing.

The ratio $Q_\text{lat}(t,t_\text{FV})/Q_\text{lat}(t-a,t_\text{FV}+a)$ is easier to calculate than $Q_\text{lat}(t,t_\text{FV})$ because the numerator and denominator actions are already very similar, differing only at two points in time. As a result, we only need one intermediate action
\ba 
\frac{Q_\text{lat}(t, t_\text{FV})}{Q_\text{lat}(t-a, t_\text{FV}+a)} = \frac{\int\mathcal{D}x \exp({-S_N})}{\int\mathcal{D}x \exp({-S_D})} = \frac{\int\mathcal{D}x \exp({-S_L(1)})}{\int\mathcal{D}x \exp({-S_L(0)})} \left(\frac{\int\mathcal{D}x \exp({-S_M(1)})}{\int\mathcal{D}x \exp({-S_M(0)})}\right)^{-1}.
\ea 

\section{Results} \label{section_results}

All the code used to generate these results is available as part of the QLattice library~\cite{Qlattice}.

\subsection{Decay rates} \label{section_results_decay_rates}
\begin{figure}
	\centering
	\includegraphics[width=\linewidth]{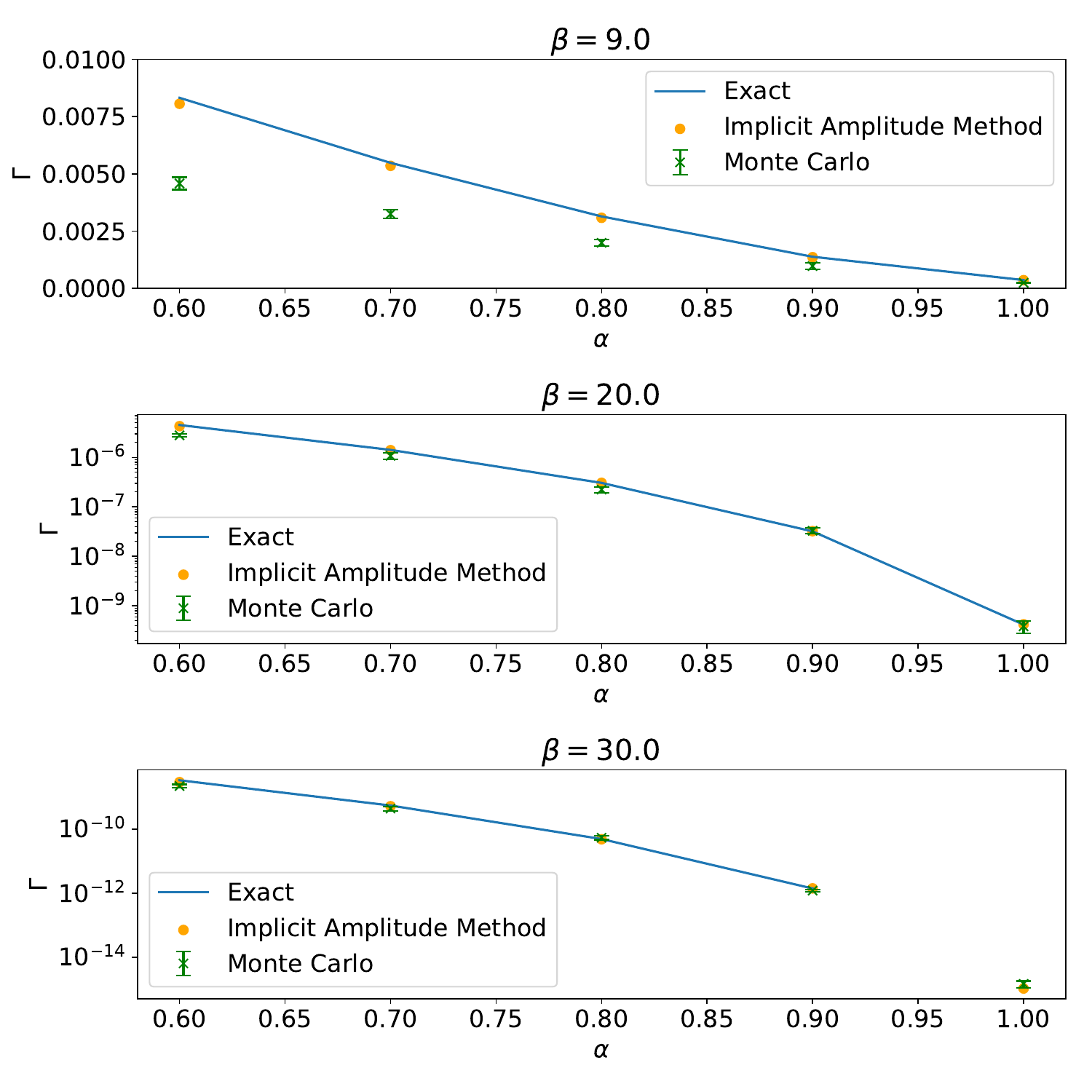}
	\caption{The exact decay rates, decay rates from the implicit decay amplitude method (Eq. \ref{eq_FGR}), and the decay rates calculated from lattice Monte Carlo are plotted for various choices of the parameters $\alpha$ and $\beta$. The errors shown are statistical only. The exact decay rates and the ``Implicit Amplitude Method'' decay rates are determined by numerically solving the Sch\"{o}dinger equation, as explained in Section \ref{section_results_decay_rates}. The difference between the Monte Carlo and the ``Implicit Amplitude Method'' results is due to error in the spectral reconstruction. Note that the last two plots use a log scale for the decay rate while the first does not. The simulation parameters and numerical results are shown in Tables \ref{tab_sim_params} and \ref{tab_sim_results}.}
	\label{decay_rate_results}
\end{figure}

Our final results for the decay rates are shown in Figure \ref{decay_rate_results} and Table \ref{tab_sim_results}. The simulation parameters used to obtain these results are given in Tables \ref{tab_sim_params} and \ref{tab_sim_results}. We show the decay rates determined by three different methods.
\begin{itemize}
	\item The exact decay rates are determined by numerically solving the Schr\"{o}dinger equation. We first use the time-dependent the Schr\"{o}dinger equation to evolve the constant wavefunction $\psi(x)=1$ in Euclidean time with $H_\text{FV}$. This projects out $|\text{FV}\rangle$. We then numerically calculate the eigenvectors $|E\rangle$ of $H$ in order to get $|\langle E|\text{FV}\rangle|^2$ as a function of $E$. Finally, we fit this spectrum near $E_\text{FV}$ with a Breit-Wigner distribution to get the decay rate. This method allows us to calculate even very small decay rates with high precision.
	\item The ``Implicit Amplitude Method'' decay rates are also determined by numerically solving the Sch\"{o}dinger equation in Euclidean time. In this case, we calculate the implicit decay amplitude 
	\ba 
	\rho(E_\text{FV}) = \langle \text{FV}|e^{-H_\text{proj}a}\delta(H_\text{TV}-E_\text{FV})e^{-H_\text{proj}a}|\text{FV}\rangle
	\ea
    and use Equation \ref{eq_FGR} (which was inspired by Fermi's golden rule) to relate $\rho(E_\text{FV})$ to the decay rate. These results incorporate systematic error from our use of the implicit decay amplitude method (Eq. \ref{eq_FGR}), from discretizing $H_\text{FV}-H\approx\frac{1}{a}\left(e^{-Ha}-e^{-H_\text{FV}a}\right)$, and from using $\epsilon\neq 0$ in defining $V_\text{proj}$ (Equation \ref{eq_def_V_proj}). We also use the same discretization of time as in our lattice results. This covers all sources of systematic error except the error from spectral reconstruction.
	\item Finally, we show the decay rate calculated using lattice Monte Carlo. In addition to the statistical errors (which are shown by the error bars in Figure \ref{decay_rate_results}), and the systematic errors present in the ``Fermi's golden rule'' results (discussed above), these results also have systematic error from the spectral reconstruction. As the figure shows, spectral reconstruction is by far the most significant source of systematic error.
\end{itemize}

\begin{table}
	\centering
	\begin{tabular}{|| c | c ||} 
		\hline
        $a$ (lattice spacing for discretizing action, Eq. \ref{eq_action}) & 0.1 \\
        \hline
        $N_t$ (number of lattice sites) & 200 \\
        \hline
        Number of trajectories in Monte Carlo simulation & 50,000 \\
		\hline
        Number of trajectories used for thermalization & 10,000 \\
		\hline
        Jackknife block size & 2,000 \\
		\hline
        $B$ (barrier strength, Eqs. \ref{eq_V_FV} and \ref{eq_V_TV}) & 100\\
        \hline
        $x_\text{bar,FV}$ (false vacuum barrier offset, Eq. \ref{eq_V_FV}) & 0.3 \\
		\hline
        $\epsilon$ (used in definition of $V_\text{proj}$, Eq. \ref{eq_def_V_proj}) & 0.01 \\
		\hline
        \hline
        \multicolumn{2}{||c||}{Parameters for intermediate ratios (see Section \ref{section_intermediate_ratios_method})} \\
        \hline\hline
        $\{P_i\}$ (see Eq. \ref{eq_S_P}) & 0, 0.6* \\
        \hline
        $\{L_i\}$ (see Eq. \ref{eq_S_L}) & 0, 0.02*, 0.1, 0.2, $0.3^\dagger$, 0.4, 0.6, 0.8 \\
        \hline
        $\{M_i\}$ (see Eq. \ref{eq_S_M}) & 0, 0.1, 0.5, 0.8 \\
        \hline
	\end{tabular}
	\caption{Simulation parameters used in Figure \ref{decay_rate_results}. *This parameter was not used for $\alpha=0.8$, $\beta=9.0$. ${}^\dagger$This parameter was only used for $\alpha=0.8$, $\beta=9.0$.}
	\label{tab_sim_params}
\end{table}

\begin{table}
	\centering
	\begin{tabular}{|| c | c | c | c | c | c | c ||} 
		\hline
        $\alpha$ & $\beta$ & $\{t\}$ & $t_\text{TV,fit}$ & $\Gamma_\text{MC}$ & $\Gamma_\text{FGR}$ & $\Gamma_\text{exact}$ \\
        \hline
		\hline
        0.6 & 9.0 & $\{2a,4a,10a\}$ & $4a$ & 0.00458(27) & 0.00806 & 0.00832 \\
        \hline
        0.7 & 9.0 & $\{2a,4a,10a,14a\}$ & $4a$ & 0.00325(19) & 0.00535 & 0.00548 \\
        \hline
        0.8 & 9.0 & $\{2a,4a,10a,14a\}$ & $4a$ & 0.00199(14) & 0.00308 & 0.00314 \\
        \hline
        0.9 & 9.0 & $\{2a,4a,10a,14a\}$ & $4a$ & 0.00097(15) & 0.00136 & 0.00138 \\
        \hline
        1.0 & 9.0 & $\{2a,4a,10a,14a,20a\}$ & $10a$ & 0.000242(28) & 0.000357 & 0.000362 \\
        \hline
        0.6 & 20.0 & $\{2a,4a,10a\}$ & $4a$ & 2.80(20)e-06 & 4.28e-6 & 4.52e-6 \\
        \hline
        0.7 & 20.0 & $\{2a,4a,10a\}$ & $4a$ & 1.08(16)e-06 & 1.40e-6 & 1.42e-6 \\
        \hline
        0.8 & 20.0 & $\{2a,4a,10a,14a\}$ & $4a$ & 2.24(32)e-07 & 3.06e-7 & 3.06e-7 \\
        \hline
        0.9 & 20.0 & $\{2a,4a,10a,14a,20a\}$ & $10a$ & 3.29(44)e-08 & 3.24e-8 & 3.21e-8 \\
        \hline
        1.0 & 20.0 & $\{2a,4a,10a,14a,20a\}$ & $10a$ & 3.9(11)e-10 & 4.25e-10 & 4.19e-10 \\
        \hline
        0.6 & 30.0 & $\{2a,4a,10a\}$ & $4a$ & 2.19(28)e-9 & 2.9e-9 & 3.32e-9 \\
        \hline
        0.7 & 30.0 & $\{2a,4a,10a\}$ & $4a$ & 4.34(66)e-10 & 5.15e-10 & 5.43e-10 \\
        \hline
        0.8 & 30.0 & $\{2a,4a,10a\}$ & $10a$ & 5.39(75)e-11 & 4.82e-11 & 4.90e-11 \\
        \hline
        0.9 & 30.0 & $\{2a,4a,10a,14a,20a\}$ & $14a$ & 1.21(11)e-12 & 1.42e-12 & 1.41e-12 \\
        \hline
        1.0 & 30.0 & $\{2a,4a,10a,14a,20a\}$ & $20a$ & 1.50(35)e-15 &  1.08e-15 & - \\
        \hline
	\end{tabular}
	\caption{Simulation parameters and results for Figure \ref{decay_rate_results}. $\alpha$ and $\beta$ are the parameters in the action (see Equation \ref{eq_action}). $\{t\}$ gives the values of $t$ at which ${Q_\text{lat}(t, t_\text{FV})}/{Q_\text{lat}(t-a, t_\text{FV}+a)}$ was calculated in order to fit the spectrum (see Section \ref{section_lattice_obs}). $t_\text{TV,fit}$ gives that value of $t$ at which $Q_\text{lat}(t, t_\text{FV})$ was calculated. $\Gamma_\text{MC}$ is the decay rate determined by Monte Carlo (errors are only statistical). $\Gamma_\text{FGR}$ and $\Gamma_\text{exact}$ are, respectively, the ``Fermi's golden rule'' and ``exact'' decay rates discussed in Section \ref{section_results_decay_rates}.}
	\label{tab_sim_results}
\end{table}

\subsection{Spectral reconstruction}
\begin{figure}
	\centering
	\includegraphics[width=\linewidth]{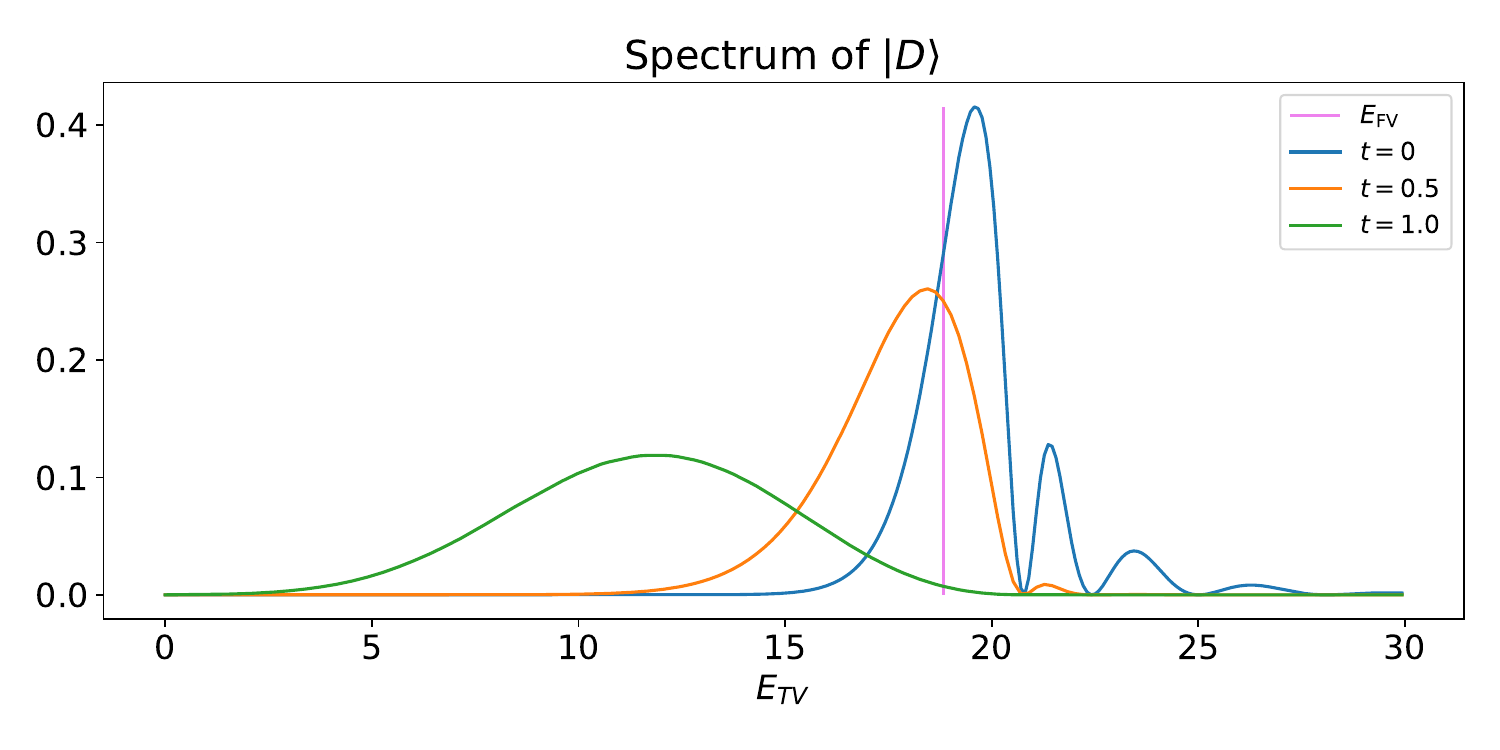}
	\caption{The energy spectrum $\rho(E)$ (defined in Equation \ref{eq_rho}) of the time-evolved difference state $e^{-H_\text{TV}t}|D\rangle = e^{-H_\text{TV}t}e^{-aH_\text{proj}}|\text{FV}\rangle$ is plotted at various times $t$ ($H_\text{proj}$ is defined in Section \ref{section_lattice_obs}). Here we set $\alpha=0.8$ and $\beta=20.0$ in the Lagrangian. Other parameters are the same as those given in Table \ref{tab_sim_params}. Since we want to fit the spectrum to a Gaussian distribution and get $\rho(E_\text{FV})$, our results will be most accurate when $t$ is long enough to suppress higher-energy components but short enough that $\rho(E_\text{FV})$ is still large.}
	\label{diff_state_spec}
\end{figure}

The main source of systematic error in our calculation is the spectral reconstruction. The spectrum of the difference state $|D\rangle$ (defined in Equation \ref{eq_rho}) is plotted in Figure \ref{diff_state_spec} after Euclidean time evolution with $e^{-(H_\text{TV}-E_\text{FV})t}$ for various choices of $t$. Note that our Gaussian ansatz works best after high-energy components are suppressed by Euclidean time evolution but before $\rho(E)$ becomes supressed at $E=E_\text{FV}$.

\begin{figure}
	\centering
	\includegraphics[width=\linewidth]{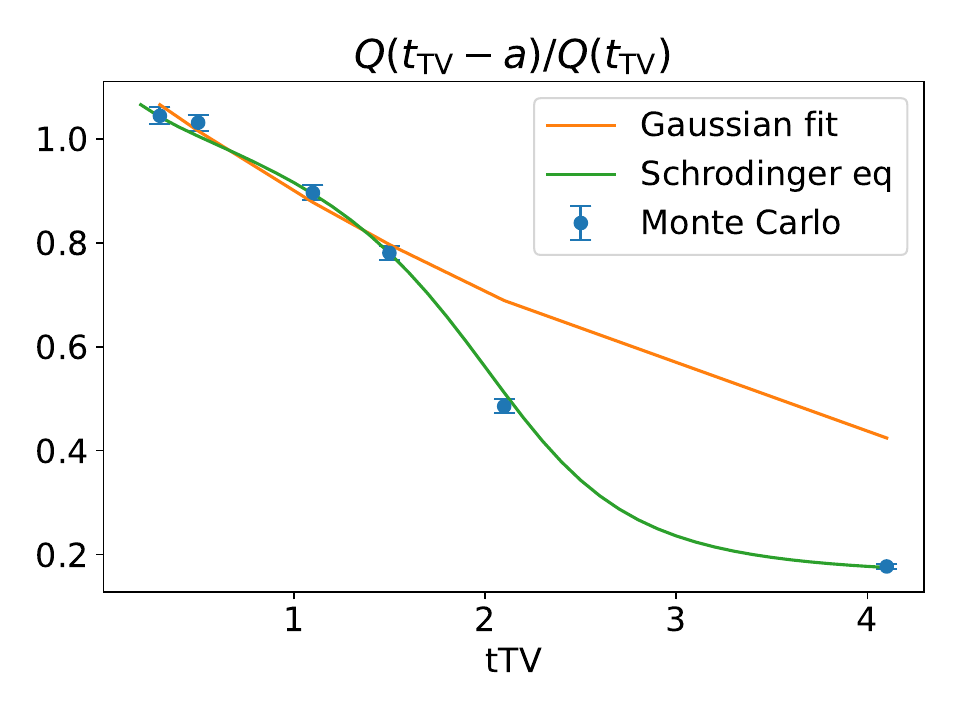}
	\caption{Using the Gaussian ansatz from Section \ref{section_inverse_prob}, we can fit the ratio ${Q_\text{lat}(t_\text{TV}, t_\text{FV})}/{Q_\text{lat}(t_\text{TV}-a, t_\text{FV}+a)}$ (discussed in Section \ref{section_lattice_obs}) as a function of $t_\text{TV}$. This fit then allows us to reconstruct the spectrum of the difference state $|D\rangle$. Here we set $\alpha=0.8$ and $\beta=20.0$ in the Lagrangian. Other parameters are the same as those given in Table \ref{tab_sim_params}. The fit is performed using only the first four Monte Carlo data points.}
	\label{fitting_spectrum}
\end{figure}

In Figure \ref{fitting_spectrum}, we show a fit to the ratio ${Q_\text{lat}(t, t_\text{FV})}/{Q_\text{lat}(t-a, t_\text{FV}+a)}$ (discussed in Section \ref{section_lattice_obs}) using the Gaussian ansatz from Section \ref{section_inverse_prob}. When the fit is very close to the actual data, we can have a high degree of confidence in our reconstruction of the spectrum. Significant departures from the fit can be caused by the higher-energy components shown in Figure \ref{diff_state_spec} (for small $t$), and by projecting out the ground state (for large $t$).

\subsection{Intermediate ratios}\label{section_intermediate_ratios_error}
\begin{figure}
	\centering
	\includegraphics[width=\linewidth]{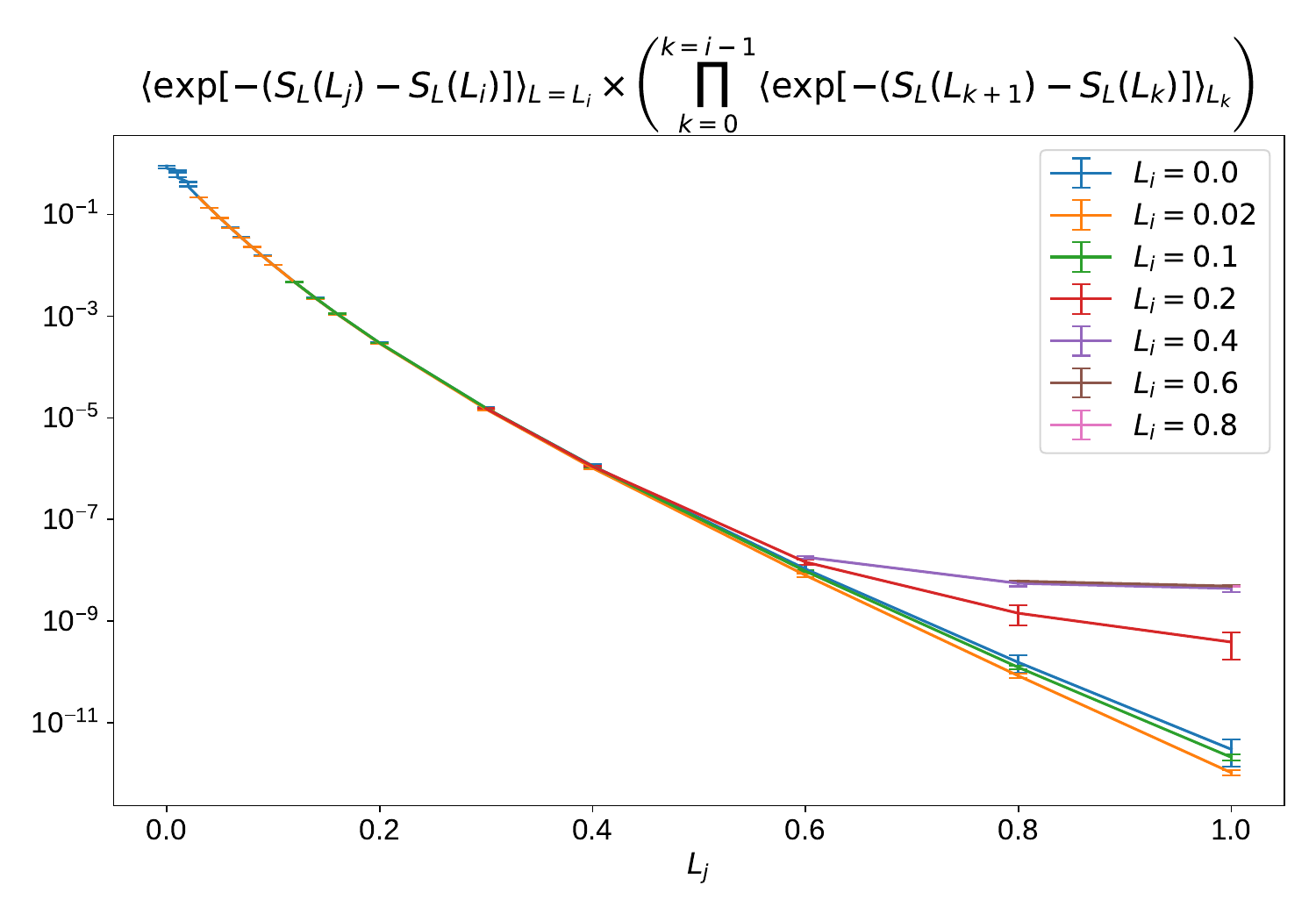}
	\caption{The ratio $\langle\exp[-(S_L(L_j)-S_L(L_i)]\rangle_{L=L_i}\times\left(\prod_{k=0}^{k=i-1}\langle\exp[-(S_L(L_{k+1})-S_L(L_k)]\rangle_{L_k}\right)$ $= {\int\mathcal{D}x \exp(-S_L(L_{j}))} / {\int\mathcal{D}x \exp({-S_L(L=0)})}$ is plotted for various choices of the final denominator action $S_L(L_i)$. The $x$-axis gives $L_j$ for the numerator action $S_L(L_j)$. Here we set $\alpha=0.8$ and $\beta=20.0$ in the Lagrangian. Other parameters are the same as those given in Table \ref{tab_sim_params}. The results from different ensembles are consistent with each other as long as the difference in the actions $S_L(L_j)-S_L(L_i)$ is not too large. We see that we will get consistent results for the final value at $L_j=1$ whether that largest value of $L_i$ that we simulate is $L_i=0.4$, $0.6,$ or $0.8$.}
	\label{intermediate_ratios}
\end{figure}

In Section \ref{section_intermediate_ratios_method}, we showed how the results from multiple ensembles can be combined to build up a ratio of the form
\ba 
\frac{\int\mathcal{D}x \exp(-S_N)}{\int\mathcal{D}x \exp({-S_D})}.
\ea 
For Figure \ref{intermediate_ratios}, we show how this works for just one family of intermediate ratios. We construct a series of intermediate ratios $\{S_L(L_i)\}$ (as defined in Equation \ref{eq_S_L}). For the ensemble generated using $L=0$, we simply plot the expectation value 
\ba 
\langle\exp[-(S_L(L_j)-S_L(0)]\rangle_{L=0} &\equiv \frac{\int\mathcal{D}x \exp[-(S_L(L_j)-S_L(0))]\exp(-S_L(0))}{\int\mathcal{D}x \exp(-S_L(0))} \\
&= \frac{\int\mathcal{D}x \exp(-S_L(L_j))}{\int\mathcal{D}x \exp(-S_L(0))}.
\ea 
Then for $L_1$, we instead plot 
$$
\langle\exp[-(S_L(L_j)-S_L(L_1)]\rangle_{L=L_1}\times\langle\exp[-(S_L(L_1)-S_L(0)]\rangle_{L=0}
$$
\ba 
= \frac{\int\mathcal{D}x \exp(-S_L(L_j)}{\int\mathcal{D}x \exp({-S_L(0)})}
\ea 
where the second factor is what we calculated on the previous ensemble. Continuing this pattern, for each subsequent ensemble, we plot 
$$
\langle\exp[-(S_L(L_j)-S_L(L_i)]\rangle_{L=L_i}\times\left(\prod_{k=0}^{k=i-1}\langle\exp[-(S_L(L_{k+1})-S_L(L_k)]\rangle_{L_k}\right)
$$
\ba
= \frac{\int\mathcal{D}x \exp(-S_L(L_{j}))}{\int\mathcal{D}x \exp({-S_L(L=0)})}
\ea
This means that, in Figure \ref{intermediate_ratios}, we plot many different ways of calculating the same ratios. From the figure, it can be seen that calculations of this same quantity on different ensembles are consistent with each other as long as the difference in the actions $S_L(L_j)-S_L(L_i)$ is not too large. We see that we will get consistent results for the final value at $L_j=1$ whether that largest value of $L_i$ that we simulate is $L_i=0.4$, $0.6,$ or $0.8$.

\section{Future work} \label{section_future_work}

\subsection{Generalization to field theory} \label{section_generalization_field_theory}
In a field theory, the false vacuum will be distinguished from the true vacuum by the value of some collective variable, such as $\bar{\phi}\equiv \int d^3x\phi(\mathbf{x},t)$ from Equation \ref{eq_bar_phi}. Let us consider how to define the false vacuum Hamiltonian for the field theory discussed in Section \ref{subsection_simple_example}. Naively, we might define, for example, the non-local Hamiltonian
\ba 
H_\text{FV,naive}\equiv 
\begin{cases}
    H & \text{for }\bar{\phi}<\phi_0 \\
    H + B(\bar{\phi} - \phi_0)^2 & \text{otherwise} \\
\end{cases}
\ea
(While non-local Hamiltonians are more difficult to simulate than local Hamiltonians, they can be handled using algorithms like Hybrid Monte Carlo.) Unfortunately, this is not a good definition of the false vacuum Hamiltonian. The problem comes from the fact that the the quantum false vacuum state can form small localized bubbles of the true vacuum state. If a bubble gets too large, it becomes stable and will continue to expand until the entire system is converted to the true vacuum state (see, for example, the discussion in~\cite{Coleman_FVD:1977}). To make sure that we do not allow any stable bubbles to form, we must choose $\phi_0$ to be far on the false vacuum side of the potential barrier. However, choosing $\phi_0$ so small may also remove non-localized fluctuations which should be included in the false vacuum state.

Instead, we can choose a false vacuum Hamiltonian that specifically targets localized fluctuations. We first define a smeared field $\phi_\text{smeared}(x)$ which smooths out the fluctuations in $\phi(x)$. Then, we can define a non-local Hamiltonian
\ba 
H_\text{FV}\equiv 
\begin{cases}
    H & \text{for }\max(\phi_\text{smeared}(x))<\phi_0 \\
    H + B (\max(\phi_\text{smeared}(x)) - \phi_0)^2 & \text{otherwise,} \\
\end{cases}
\ea
where $\max(\phi_\text{smeared}(x))$ returns the maximum value of $\phi_\text{smeared}(x)$ as $x$ is varied. 

Note that instead of calculating how long the field takes to be completely converted to the true vacuum state, we can simply calculate how long it takes for a stable bubble to form. Once a large enough bubble forms, we know that the rest of the field will eventually transition to the true vacuum.

\subsection{Continuous intermediate ratios}
Instead of performing many Monte Carlo simulations for the intermediate ratios of Section \ref{section_intermediate_ratios_method}, we can write 
\ba
\frac{\int\mathcal{D}x \exp(-S_{\lambda_n})}{\int\mathcal{D}x \exp({-S_{\lambda_0}})} &= \prod_i\frac{\int\mathcal{D}x \exp\left(-(S_{\lambda_{i+1}}-S_{\lambda_i})\right)\exp(-S_{\lambda_i})}{\int\mathcal{D}x \exp(-S_{\lambda_i})}
\\
&=\exp\left[\sum \log\left(\frac{\int\mathcal{D}x \exp\left(-(S_{\lambda_{i+1}}-S_{\lambda_i})\right)\exp(-S_{\lambda_i})}{\int\mathcal{D}x \exp(-S_{\lambda_i})}\right)\right]
\\
&=\exp\left[\int_{\lambda_0}^{\lambda_1} d\lambda \log\left(\frac{\int\mathcal{D}x \exp(-\frac{d}{d\lambda}S_\lambda)\exp({-S_\lambda})}{\int\mathcal{D}x \exp(-S_\lambda)}\right)\right].
\ea
The integral over $\lambda$ can be performed by the Monte Carlo method, or with some quadrature rule.

\subsection{Multi-action Monte Carlo}
When calculating $Q_\text{lat}(t,t_\text{FV})/Q_\text{lat}(t-a,t_\text{FV}+a)$, instead of performing two Monte Carlo simulations as in Section \ref{section_intermediate_ratios_calc_obs} to calculate the ratio as
\ba
\frac{\int\mathcal{D}x \exp({-S_N})}{\int\mathcal{D}x \exp({-S_D})} = \frac{\int\mathcal{D}x \exp({-S_L(1)})}{\int\mathcal{D}x \exp({-S_L(0)})} \left(\frac{\int\mathcal{D}x \exp({-S_M(1)})}{\int\mathcal{D}x \exp({-S_M(0)})}\right)^{-1},
\ea
we can combine the two intermediate actions into one big Monte Carlo simulation: 
$$\frac{\int\mathcal{D}x \exp({-S_N})}{\int\mathcal{D}x \exp({-S_D})} = \frac{\int\mathcal{D}x \exp({-S_L(1)})}{\int\mathcal{D}x \left[\exp({-S_L(0)}) + \exp({-S_M(0)})\right]}$$
\ba 
\times \left(\prod_k\frac{\int\mathcal{D}x \exp({-S_M(1)})}{\int\mathcal{D}x \left[\exp({-S_L(0)}) + \exp({-S_M(0)})\right]}\right)^{-1}.
\ea
Only one Monte Carlo simulation is needed because both observables are over a common denominator.

\section{Conclusion}
We have developed a method for calculating false vacuum decay rates using Euclidean-time lattice Monte Carlo simulations. In Sections \ref{section_FGR}, \ref{section_FGR_derivation}, and \ref{section_lattice_obs}, we related the decay rate to a Euclidean-time observable through the implicit decay amplitude method (inspired by Fermi's golden rule). In Section \ref{section_inverse_prob}, we discussed a simple method for spectral reconstruction. Finally, to measure our observables, we developed a new sampling method in Section \ref{section_intermediate_ratios_method} which eliminated problems with ergodicity and signal suppression at the cost of requiring multiple Monte Carlo simulations. In Section \ref{section_results}, we presented our results for a simple one-dimensional quantum tunneling problem. In Section \ref{section_generalization_field_theory}, we discussed the differences between this tunneling problem and false vacuum decay in a full quantum field theory.

Our Monte Carlo results for the decay rates can differ from the exact results by a factor less than $2$. The difference is mainly due to the error from spectral reconstruction. This error could presumably be reduced by using more sophisticated methods for spectral reconstruction. Many methods are available~\cite{Hansen:2019idp,Bailas:2020qmv,Bergamaschi:2023xzx,Fields:2025glg,Rietsch:1977,Asakawa:2000tr,Burnier:2013nla,Backus:1968svk,Backus:1970,Hansen:2017mnd,Abbott:2025snz} which could perform better than our simple gaussian ansatz and allow to estimation of systematic error. It should be noted, however, that we may find the spectral reconstruction easier in field theory than in single-particle quantum mechanics. In a field theory, because there are more degrees of freedom, the density of states increases much more rapidly with energy. As a result, after we evolve our initial state with $e^{-(H_\text{TV}-E_\text{FV})t}$ to suppress higher-energy states, we will likely end up with a state which has a very narrowly-peaked energy spectrum. This is analogous to the way the canonical ensemble often can be approximated by the microcanonical ensemble. The Boltzmann factor $e^{-\beta E}$ plays the role of Euclidean time evolution suppressing higher-energy states, while density of states rapidly increases with energy so that the canonical ensemble is mostly made up of states peaked around a single energy.

Another method for calculating false vacuum decay rates (using Euclidean-time lattice Monte Carlo) was proposed in Ref.~\cite{Shen:2022}. This method was tested on the same quantum tunneling problem (using the action from Equation \ref{eq_action}), and so it is interesting to compare our results. Their results for $\beta=9$ are more accurate than ours (except perhaps for $\alpha=0.9$). They do not have results for $\beta=20$ because their method cannot be used for very small decay rates. They do propose an alternative method for small decay rates that works on small lattice volumes, and it is possible that their main method could be extended to smaller decay rates using more sophisticated sampling methods in the future. The sources of systematic error between the two methods are very different. In their method, they solve the suppression problem by rejecting any configurations with actions below a certain cut. They estimate, using the semi-classical approximation, that this can result in their decay rates being off by a factor of order 1. They also use the semi-classical approximation to estimate the probability flow velocity out of the false vacuum at the classical turning point. Our method does not rely on semi-classical approximations, but it does suffer from systematic error due to analytic continuation.

In the future, we hope to test our method on false vacuum decay in a full field theory. Other future directions of research could include introducing more sophisticated spectral reconstruction with estimates of systematic errors and attempting to generalize this method to finite temperature.

\section{Acknowledgments}

We thank our RBC and UKQCD collaborators, and especially Tom Blum, for helpful discussions and and critical software and hardware support. L.J. and J.S. acknowledge the support of DOE Office of Science Early Career Award DE-SC0021147, DOE grant DE-SC0010339 and DE-SC0026314.
We developed the computational code used for this work based on \href{https://github.com/jinluchang/Qlattice}{Qlattice}~\cite{Qlattice}. The research reported in this work made use of computing facilities of the USQCD Collaboration, which are funded by the Office of Science
of the U.S. Department of Energy. This work was supported by resources provided by the Scientific Data and Computing Center (SDCC), a component of the Computational Science Initiative (CSI) at Brookhaven National Laboratory (BNL) and a DOE Office of Science User Facility supported by the Office of Science of the U.S. Department of Energy. Part of the computational work for this project was conducted using resources provided by the Storrs High-Performance Computing (HPC) cluster. We extend our gratitude to the UConn Storrs HPC and its team for their resources and support, which aided in achieving these results.

\appendix

\section{Fermi's golden rule for false vacuum systems} \label{appendix_FGR_formal}
We can formally define $H_0$ and $H_\text{tr}$ such that $H=H_0+H_\text{tr}$, where $H_0$ does not allow false vacuum decay, and $H_\text{tr}$ is a small perturbation (in that its matrix elements are small for relevant states). To begin, we need to formally divide our Hilbert space into a false vacuum and true vacuum region. Let $\tilde{P}_\text{FV}$ be the ``harsh'' projection operator, given in the position basis by 
\ba
\tilde{P}_\text{FV}(x)=\begin{cases}
	1 & x<x_\text{threshold} \\
	0 & x_\text{threshold}\le x
\end{cases}
\ea
and define $\tilde{P}_\text{TV}\equiv 1-\tilde{P}_\text{FV}$. Naively, these two projection operators accomplish our goal. However, we immediately run into problems because $\tilde{P}_\text{TV}\psi$ and $\tilde{P}_\text{FV}\psi$ have infinite energy for any state $\psi$ with $\psi(x_\text{div})\neq 0$. Even if we fix this problem by making a smoothed version of $\tilde{P}_\text{FV}$ and $\tilde{P}_\text{TV}$, the resulting division of true and false vacuum would be rather arbitrary. The false vacuum and true vacuum are only well-separated for low-energy states. For sufficiently high energy states, the barrier between the true vacuum and false vacuum is negligible.

With this understanding, we define $P_\text{Low}$ as the operator that projects onto the space spanned by energy eigenstates with energies less than some threshold energy:
\ba 
P_\text{Low} = \sum_{E_n<E_\text{threshold}}|n\rangle\langle n|.
\ea
The threshold energy should be less than the height of the potential barrier, but large enough to include the band around the lowest resonant energy. We can also define a projection operator onto high-energy states by $P_\text{High}=1-P_\text{Low}$. Then we moderate our ``harsh'' projection operator $\tilde{P}_\text{FV}$ by first projecting onto the low-energy subspace
\ba 
F = P_\text{Low}\tilde{P}_\text{FV}P_\text{Low}.
\ea
$F$ is a Hermitian operator, and can therefore be diagonalized (it is also compact, unlike $\tilde{P}_\text{FV}$). The eigenvalues of $F$ all lie between 0 and 1. An eigenstate with eigenvalue close to 1 is a low-energy state localized mostly in the false vacuum, and a state with eigenvalue close to 0 has its low-energy components localized mostly in the true vacuum. Note also that states with low-energy components that have significant probability in both the false and true vacuum cannot be eigenstates of $F$. Therefore, we divide the eigenvectors of $F$ into two categories based on whether the corresponding eigenvalue is less than 0.5 or not (the exact choice of 0.5 is not important since most eigenvalues are close to either 0 or 1). Then we can finally define a ``moderate'' projection operator $P_\text{FV}$ that projects onto the space spanned by eigenvectors of $F$ with eigenvalues greater than 0.5,
\ba 
P_\text{FV}\equiv \theta\left(P_\text{Low}\tilde{P}_\text{FV}P_\text{Low}-\frac{1}{2}\right).
\ea 
We define the ``moderate'' true vacuum projection operator as 
\ba 
P_\text{TV}=P_\text{Low}-P_\text{FV}.
\ea 
Note that $P_\text{FV}$ already projects onto the low-energy subspace, and that $P_\text{FV}+P_\text{TV}=P_\text{Low}$.


If we define the high-energy projector $P_\text{High}=1-P_\text{Low}$, we can now write our Hamiltonian as
\ba 
H = (P_\text{TV}+P_\text{FV}+P_\text{High})H(P_\text{TV}+P_\text{FV}+P_\text{High})=H_0+H_\text{tr},
\ea 
where
\ba 
H_0 \equiv P_\text{TV}HP_\text{TV} + P_\text{FV}HP_\text{FV} + P_\text{High}HP_\text{High}\quad\text{and}\quad
H_\text{tr} \equiv P_\text{TV}HP_\text{FV}+P_\text{FV}HP_\text{TV}.
\ea 
Note that acting with $H$ cannot cause a transition between the low-energy and high-energy subspaces, and so we do not need to include terms which mix high and low energy projectors, such as $P_\text{FV}HP_\text{High}$. With this division, we see now that $H_\text{tr}$ is the part of the Hamiltonian which is responsible for transitions of low-energy states from the false to the true vacuum, and vice versa. Such transitions are not allowed under evolution with $H_0$.

Consider the action of $P_\text{TV}HP_\text{FV}$ on a state $\psi$. Acting with $P_\text{FV}$ will project out a low-energy state that is mainly localized in the false vacuum. Now if we were to evolve $P_\text{FV}\psi$ with $H$, it would leak out slowly. Therefore, the norm of $P_\text{TV}HP_\text{FV}\psi$ is suppressed compared to the norm of $HP_\text{FV}\psi$. By the same reasoning, the norm of $P_\text{FV}HP_\text{TV}\psi$ is suppressed. Therefore, we can therefore treat $H_\text{tr}$ as a small perturbation from $H_0$.

Let $|\text{FV}\rangle$ be the false vacuum state. By Fermi's golden rule, the transition rate from this low-energy state to a low-energy true vacuum state is
\ba 
\Gamma = 2\pi \sum_{n} |\langle n|H_\text{tr}|\text{FV}\rangle|^2\delta(E_n-E_\text{FV})=2\pi\langle \text{FV}|H_\text{tr} \delta(H_0-E_\text{FV})H_\text{tr}|\text{FV}\rangle,
\ea 
where $E_\text{FV}\equiv \langle\text{FV}|H_0|\text{FV}\rangle$ is the false vacuum energy.

\bibliography{false_vacuum_decay}

@misc{Qlattice,
	title = {Qlattice},
	publisher = {GitHub},
	year = {2025},
	url = {https://github.com/jinluchang/Qlattice},
	howpublished = {\url{https://github.com/jinluchang/Qlattice}},
}

@article{Langer:1967,
	author = "Langer, J. S.",
	title = "{Theory of the condensation point}",
	doi = "10.1016/0003-4916(67)90200-X",
	journal = "Annals Phys.",
	volume = "41",
	pages = "108--157",
	year = "1967"
}

@article{Coleman_FVD:1977,
	author = "Coleman, Sidney R.",
	title = "{The Fate of the False Vacuum. 1. Semiclassical Theory}",
	reportNumber = "HUTP-77-A004",
	doi = "10.1103/PhysRevD.16.1248",
	journal = "Phys. Rev. D",
	volume = "15",
	pages = "2929--2936",
	year = "1977",
	note = "[Erratum: Phys.Rev.D 16, 1248 (1977)]"
}

@article{Callan_FVD:1977,
	author = "Callan, Jr., Curtis G. and Coleman, Sidney R.",
	title = "{The Fate of the False Vacuum. 2. First Quantum Corrections}",
	reportNumber = "HUTP-77-A032",
	doi = "10.1103/PhysRevD.16.1762",
	journal = "Phys. Rev. D",
	volume = "16",
	pages = "1762--1768",
	year = "1977"
}

@article{Linde_finite_T:1980,
	author = "Linde, Andrei D.",
	title = "{Fate of the False Vacuum at Finite Temperature: Theory and Applications}",
	reportNumber = "LEBEDEV-80-92",
	doi = "10.1016/0370-2693(81)90281-1",
	journal = "Phys. Lett. B",
	volume = "100",
	pages = "37--40",
	year = "1981"
}

@article{Linde_finite_T:1981,
	author = "Linde, Andrei D.",
	title = "{Decay of the False Vacuum at Finite Temperature}",
	reportNumber = "LEBEDEV-81-265",
	doi = "10.1016/0550-3213(83)90072-X",
	journal = "Nucl. Phys. B",
	volume = "216",
	pages = "421",
	year = "1983",
	note = "[Erratum: Nucl.Phys.B 223, 544 (1983)]"
}

@article{Shen:2022,
	author = "Shen, Jiayu and Draper, Patrick and El-Khadra, Aida X.",
	title = "{Vacuum decay and Euclidean lattice Monte~Carlo}",
	eprint = "2210.05925",
	archivePrefix = "arXiv",
	primaryClass = "hep-lat",
	doi = "10.1103/PhysRevD.107.094506",
	journal = "Phys. Rev. D",
	volume = "107",
	number = "9",
	pages = "094506",
	year = "2023"
}

@article{Andreassen:2017,
	author = "Andreassen, Anders and Farhi, David and Frost, William and Schwartz, Matthew D.",
	title = "{Precision decay rate calculations in quantum field theory}",
	eprint = "1604.06090",
	archivePrefix = "arXiv",
	primaryClass = "hep-th",
	doi = "10.1103/PhysRevD.95.085011",
	journal = "Phys. Rev. D",
	volume = "95",
	number = "8",
	pages = "085011",
	year = "2017"
}

@article{Bezuglov_two_loop_scalar:2018,
	author = "Bezuglov, M. A. and Onishchenko, A. I.",
	title = "{Two-loop corrections to false vacuum decay in scalar field theory}",
	eprint = "1805.06482",
	archivePrefix = "arXiv",
	primaryClass = "hep-ph",
	doi = "10.1016/j.physletb.2018.11.005",
	journal = "Phys. Lett. B",
	volume = "788",
	pages = "122--130",
	year = "2019"
}

@article{Dunne:2005,
	author = "Dunne, Gerald V. and Min, Hyunsoo",
	title = "{Beyond the thin-wall approximation: Precise numerical computation of prefactors in false vacuum decay}",
	eprint = "hep-th/0511156",
	archivePrefix = "arXiv",
	doi = "10.1103/PhysRevD.72.125004",
	journal = "Phys. Rev. D",
	volume = "72",
	pages = "125004",
	year = "2005"
}

@article{Weinberg_FVD_by_rad:1992,
	author = "Weinberg, Erick J.",
	title = "{Vacuum decay in theories with symmetry breaking by radiative corrections}",
	eprint = "hep-ph/9211314",
	archivePrefix = "arXiv",
	reportNumber = "CU-TP-577, NSF-ITP-93-48",
	doi = "10.1103/PhysRevD.47.4614",
	journal = "Phys. Rev. D",
	volume = "47",
	pages = "4614--4627",
	year = "1993"
}

@article{Croon_nonpert:2021,
	author = "Croon, Djuna and Hall, Eleanor and Murayama, Hitoshi",
	title = "{Non-perturbative methods for false vacuum decay}",
	eprint = "2104.10687",
	archivePrefix = "arXiv",
	primaryClass = "hep-th",
    doi = "10.48550/arXiv.2104.10687",
    url={https://arxiv.org/abs/2104.10687},
	reportNumber = "IPPP/20/95",
	month = "4",
	year = "2021"
}

@article{Bai_flow_based:2024,
	author = "Bai, Yang and Chen, Ting-Kuo",
	title = "{Flow-based nonperturbative simulation of first-order phase transitions}",
	eprint = "2404.18323",
	archivePrefix = "arXiv",
	primaryClass = "hep-lat",
	doi = "10.1007/JHEP10(2024)198",
	journal = "JHEP",
	volume = "10",
	pages = "198",
	year = "2024"
}

@article{Braden:2018,
	author = "Braden, Jonathan and Johnson, Matthew C. and Peiris, Hiranya V. and Pontzen, Andrew and Weinfurtner, Silke",
	title = "{New Semiclassical Picture of Vacuum Decay}",
	eprint = "1806.06069",
	archivePrefix = "arXiv",
	primaryClass = "hep-th",
	doi = "10.1103/PhysRevLett.123.031601",
	journal = "Phys. Rev. Lett.",
	volume = "123",
	number = "3",
	pages = "031601",
	year = "2019",
	note = "[Erratum: Phys.Rev.Lett. 129, 059901 (2022)]"
}

@article{Hertzberg_app_of_Braden_method:2019,
	author = "Hertzberg, Mark P. and Yamada, Masaki",
	title = "{Vacuum Decay in Real Time and Imaginary Time Formalisms}",
	eprint = "1904.08565",
	archivePrefix = "arXiv",
	primaryClass = "hep-th",
	doi = "10.1103/PhysRevD.100.016011",
	journal = "Phys. Rev. D",
	volume = "100",
	number = "1",
	pages = "016011",
	year = "2019"
}

@article{Wang:2025ooq,
	author = "Wang, Haiyang and Qin, Renhui and Bian, Ligong",
	title = "{Numerical simulation of the false vacuum decay at finite temperature}",
	eprint = "2506.18334",
	archivePrefix = "arXiv",
	primaryClass = "hep-th",
	month = "6",
	year = "2025"
}

@article{Andreassen_SM_lifetime:2017,
	author = "Andreassen, Anders and Frost, William and Schwartz, Matthew D.",
	title = "{Scale Invariant Instantons and the Complete Lifetime of the Standard Model}",
	eprint = "1707.08124",
	archivePrefix = "arXiv",
	primaryClass = "hep-ph",
	doi = "10.1103/PhysRevD.97.056006",
	journal = "Phys. Rev. D",
	volume = "97",
	number = "5",
	pages = "056006",
	year = "2018"
}

@article{Chigusa_SM_decay:2018,
	author = "Chigusa, So and Moroi, Takeo and Shoji, Yutaro",
	title = "{Decay Rate of Electroweak Vacuum in the Standard Model and Beyond}",
	eprint = "1803.03902",
	archivePrefix = "arXiv",
	primaryClass = "hep-ph",
	reportNumber = "UT-18-04",
	doi = "10.1103/PhysRevD.97.116012",
	journal = "Phys. Rev. D",
	volume = "97",
	number = "11",
	pages = "116012",
	year = "2018"
}

@article{Xu_meta_heavy_ion:2023,
author = "Xu, Mingmei and Wu, Yuanfang",
title = "{The Metastable State and the Finite-Size Effect of the First-Order Phase Transition}",
doi = "10.3390/sym15020510",
journal = "Symmetry",
volume = "15",
number = "2",
pages = "510",
year = "2023"
}

@article{Weir_GW_EW_review:2017,
author = "Weir, David J.",
title = "{Gravitational waves from a first order electroweak phase transition: a brief review}",
eprint = "1705.01783",
archivePrefix = "arXiv",
primaryClass = "hep-ph",
reportNumber = "HIP-2017-06-TH, HIP-2017-06/TH",
doi = "10.1098/rsta.2017.0126",
journal = "Phil. Trans. Roy. Soc. Lond. A",
volume = "376",
number = "2114",
pages = "20170126",
year = "2018",
note = "[Erratum: Phil.Trans.Roy.Soc.Lond.A 381, 20230212 (2023)]"
}

@article{Hindmarsh_GW_EW_lecture_notes:2020,
	author = {Hindmarsh, Mark B. and L\"uben, Marvin and Lumma, Johannes and Pauly, Martin},
	title = "{Phase transitions in the early universe}",
	eprint = "2008.09136",
	archivePrefix = "arXiv",
	primaryClass = "astro-ph.CO",
	reportNumber = "MPP-2020-163, HIP-2020-27/TH",
	doi = "10.21468/SciPostPhysLectNotes.24",
	journal = "SciPost Phys. Lect. Notes",
	volume = "24",
	pages = "1",
	year = "2021"
}

@article{Bodeker_baryogenesis_review:2020,
	author = "Bodeker, Dietrich and Buchmuller, Wilfried",
	title = "{Baryogenesis from the weak scale to the grand unification scale}",
	eprint = "2009.07294",
	archivePrefix = "arXiv",
	primaryClass = "hep-ph",
	reportNumber = "DESY 20-141, DESY-20-141",
	doi = "10.1103/RevModPhys.93.035004",
	journal = "Rev. Mod. Phys.",
	volume = "93",
	number = "3",
	pages = "035004",
	year = "2021"
}

@article{Moore_EW_bubble:2000,
	author = "Moore, Guy D. and Rummukainen, Kari",
	title = "{Electroweak bubble nucleation, nonperturbatively}",
	eprint = "hep-ph/0009132",
	archivePrefix = "arXiv",
	reportNumber = "UW-PT-00-01, NORDITA-2000-79-HE",
	doi = "10.1103/PhysRevD.63.045002",
	journal = "Phys. Rev. D",
	volume = "63",
	pages = "045002",
	year = "2001"
}

@article{Gould_EW_bubble:2022,
	author = {Gould, Oliver and G\"uyer, Sinan and Rummukainen, Kari},
	title = "{First-order electroweak phase transitions: A nonperturbative update}",
	eprint = "2205.07238",
	archivePrefix = "arXiv",
	primaryClass = "hep-lat",
	reportNumber = "HIP-2022-10/TH",
	doi = "10.1103/PhysRevD.106.114507",
	journal = "Phys. Rev. D",
	volume = "106",
	number = "11",
	pages = "114507",
	year = "2022",
	note = "[Erratum: Phys.Rev.D 110, 119903 (2024)]"
}

@article{Ai_Sch_effect:2020,
	author = "Ai, Wen-Yuan and Drewes, Marco",
	title = "{Schwinger effect and false vacuum decay as quantum-mechanical tunneling of a relativistic particle}",
	eprint = "2005.14163",
	archivePrefix = "arXiv",
	primaryClass = "hep-th",
	reportNumber = "CP3-20-24",
	doi = "10.1103/PhysRevD.102.076015",
	journal = "Phys. Rev. D",
	volume = "102",
	number = "7",
	pages = "076015",
	year = "2020"
}

@article{Rietsch:1977,
    author = "Rietsch, E.",
    title = "{The Maximum Entropy Approach to Inverse Problems}",
    journal = "Journal of Geophysics",
    volume = "42",
    pages = "489--506",
    year = "1977"
}

@article{Asakawa:2000tr,
    author = "Asakawa, M. and Hatsuda, T. and Nakahara, Y.",
    title = "{Maximum entropy analysis of the spectral functions in lattice QCD}",
    eprint = "hep-lat/0011040",
    archivePrefix = "arXiv",
    reportNumber = "DPNU-00-39",
    doi = "10.1016/S0146-6410(01)00150-8",
    journal = "Prog. Part. Nucl. Phys.",
    volume = "46",
    pages = "459--508",
    year = "2001"
}

@article{Burnier:2013nla,
    author = "Burnier, Yannis and Rothkopf, Alexander",
    title = "{Bayesian Approach to Spectral Function Reconstruction for Euclidean Quantum Field Theories}",
    eprint = "1307.6106",
    archivePrefix = "arXiv",
    primaryClass = "hep-lat",
    doi = "10.1103/PhysRevLett.111.182003",
    journal = "Phys. Rev. Lett.",
    volume = "111",
    pages = "182003",
    year = "2013"
}

@article{Backus:1968svk,
    author = "Backus, George and Gilbert, Freeman",
    title = "{The Resolving Power of Gross Earth Data}",
    doi = "10.1111/j.1365-246X.1968.tb00216.x",
    journal = "Geophys. J. Int.",
    volume = "16",
    number = "2",
    pages = "169--205",
    year = "1968"
}

@article{Backus:1970,
    author = {Backus, G. and Gilbert, F.},
    title = {Uniqueness in the inversion of inaccurate gross Earth data},
    journal = {Philosophical Transactions of the Royal Society of London, Series A: Mathematical and Physical Sciences},
    volume = {266},
    number = {1173},
    pages = {123-192},
    year = {1970},
    month = {03},
    issn = {0080-4614},
    doi = {10.1098/rsta.1970.0005},
    url = {https://doi.org/10.1098/rsta.1970.0005},
    eprint = {https://royalsocietypublishing.org/rsta/article-pdf/266/1173/123/268003/rsta.1970.0005.pdf},
}

@article{Hansen:2017mnd,
    author = "Hansen, Maxwell T. and Meyer, Harvey B. and Robaina, Daniel",
    title = "{From deep inelastic scattering to heavy-flavor semileptonic decays: Total rates into multihadron final states from lattice QCD}",
    eprint = "1704.08993",
    archivePrefix = "arXiv",
    primaryClass = "hep-lat",
    doi = "10.1103/PhysRevD.96.094513",
    journal = "Phys. Rev. D",
    volume = "96",
    number = "9",
    pages = "094513",
    year = "2017"
}

@article{Hansen:2019idp,
    author = "Hansen, Martin and Lupo, Alessandro and Tantalo, Nazario",
    title = "{Extraction of spectral densities from lattice correlators}",
    eprint = "1903.06476",
    archivePrefix = "arXiv",
    primaryClass = "hep-lat",
    doi = "10.1103/PhysRevD.99.094508",
    journal = "Phys. Rev. D",
    volume = "99",
    number = "9",
    pages = "094508",
    year = "2019"
}

@article{Bailas:2020qmv,
    author = "Bailas, Gabriela and Hashimoto, Shoji and Ishikawa, Tsutomu",
    title = "{Reconstruction of smeared spectral function from Euclidean correlation functions}",
    eprint = "2001.11779",
    archivePrefix = "arXiv",
    primaryClass = "hep-lat",
    reportNumber = "KEK-CP-374",
    doi = "10.1093/ptep/ptaa044",
    journal = "PTEP",
    volume = "2020",
    number = "4",
    pages = "043B07",
    year = "2020"
}

@article{Bergamaschi:2023xzx,
	author = "Bergamaschi, Thomas and Jay, William I. and Oare, Patrick R.",
	title = "{Hadronic structure, conformal maps, and analytic continuation}",
	eprint = "2305.16190",
	archivePrefix = "arXiv",
	primaryClass = "hep-lat",
	reportNumber = "MIT-CTP/5563",
	doi = "10.1103/PhysRevD.108.074516",
	journal = "Phys. Rev. D",
	volume = "108",
	number = "7",
	pages = "074516",
	year = "2023"
}

@article{Fields:2025glg,
	author = "Fields, Sarah and Christ, Norman",
	title = "{Nevanlinna-Pick interpolation from uncertain data}",
	eprint = "2510.12136",
	archivePrefix = "arXiv",
	primaryClass = "hep-lat",
	month = "10",
	year = "2025"
}

@article{Abbott:2025snz,
    author = "Abbott, Ryan and Jay, William I. and Oare, Patrick R.",
    title = "{Moment problems and bounds for matrix-valued smeared spectral functions}",
    eprint = "2508.01377",
    archivePrefix = "arXiv",
    primaryClass = "hep-lat",
    reportNumber = "MIT-CTP/5896",
    month = "8",
    year = "2025"
}

@article{Blum:2024hyr,
    author = "Blum, Thomas and Jay, William I. and Jin, Luchang and Kronfeld, Andreas S. and Stewart, Douglas Byron Allen",
    title = "{Toward inclusive observables with staggered quarks: the smeared $R${\textasciitilde}ratio}",
    eprint = "2411.14300",
    archivePrefix = "arXiv",
    primaryClass = "hep-lat",
    reportNumber = "FERMILAB-CONF-24-0836-T, MIT-CTP/5409",
    doi = "10.22323/1.466.0126",
    journal = "PoS",
    volume = "LATTICE2024",
    pages = "126",
    year = "2025"
}

@article{ExtendedTwistedMass:2025tpc,
    author = "Margari, Francesca and others",
    collaboration = "Extended Twisted Mass",
    title = "{Smeared $R$-ratio in isospin symmetric QCD with Low Mode Averaging}",
    eprint = "2502.03187",
    archivePrefix = "arXiv",
    primaryClass = "hep-lat",
    doi = "10.22323/1.466.0446",
    journal = "PoS",
    volume = "LATTICE2024",
    pages = "446",
    year = "2025"
}

\end{document}